\documentclass[useAMS, usenatbib]{mnras}
\usepackage{graphicx}
\usepackage{amssymb}
\usepackage{booktabs}
\usepackage{amsmath}
\newcommand\ApJ{ApJ}
\newcommand\MNRAS{MNRAS}

\newcommand\ApJLett{ApJL}
\newcommand\PASJ{PASJ}

\newcommand\AnA{A\&A}

\def\alf{Alfv\'en\,}
\def\alfc{Alfv\'enic\,}
\def\bq{\begin{equation}}
\def\eq{\end{equation}}
\def\ee #1 {\times 10^{#1}}
\def\ut #1 #2 { \, \rmn{#1}^{#2}}
\def\u #1 { \, \rmn{#1}}

\def\persec {\, \hbox{s}^{-1}}
\def\cc {\,\rmn{cm}^{-3}}
\def\persqcm {\, \hbox{cm}^{-2}}

\let\grad=\nabla
\newcommand\cross{\bmath{\times}}
\newcommand\etaO{\eta_\mathrm{O}}
\newcommand\etaA{\eta_\mathrm{A}}
\newcommand\etaH{\eta_\mathrm{H}}
\newcommand\etaP{\eta_\mathrm{P}}

\newcommand\bm{\bmath}
\def\curl{{\grad \cross}}
\def\div #1 {\grad \cdot #1}

\def\del{\partial}

\def\omA{\omega_A}
\def\omD{\omega_{AD}}

\def\k{\bmath{k}}
\def\kh{\hat{\k}}
\def\kz{\hat{k}_z}
\def\kx{\hat{k}_x}
\def\bb{\bm{b}}
\def\bz{b_z}
\def\by2{b_y^2}
\def\bz2{b_z^2}

\def\v{\bmath{v}}

\def\dvx{\delta v_x}
\def\dvy{\delta v_y}
\def\dvz{\delta v_z}

\def\v{\bmath{v}}

\def\vi{\bmath{v}_i}

\def\ve{\bmath{v}_e}
\def\vi{\bmath{v}_i}

\def\vn{\bmath{v}_n}

\def\B{\bmath{B}}
\def\J{\bmath{J}}

\def\dBxh{\delta b_x}

\def\dByh{\delta b_y}

\def\dBzh{\delta b_z}
\def\dbb{\delta \bb}
\def\va{v_A}
\def\Bh{\bb}
\def\vx{v_{x}}
\def\vy{v_{y}}
\def\vz{v_{z}}
\def\vp{{v_0}^{\prime}}

\def\dBh{\bmath{\delta\bb}}

\newcommand{\delt} [1] {\frac{\partial #1}{\partial t}}

\newcommand{\tenq}[1]{\hbox{\oalign{$\bm{#1}$\crcr\hidewidth$\scriptscriptstyle\bm{\approx}$\hidewidth}}}
\title{Viscous Heating and Instabilities in the Partially Ionized Solar Atmosphere}
\author[B.P.Pandey and Mark Wardle]
        {B.P. Pandey and Mark Wardle \\
{School of Mathematical and Physical Sciences, Macquarie University, Sydney, NSW 2109, Australia}}
\date{\today}
\pagerange{\pageref{firstpage}--\pageref{lastpage}}
\pubyear{2021}
\begin{document}
\maketitle
\label{firstpage}
\begin{abstract}
In weak magnetic fields ($\lesssim 50 \,\mbox{G}$), parallel and perpendicular viscosities, mainly from neutrals, may exceed magnetic diffusivities (Ohm, Hall, ambipolar) in the middle and upper chromosphere. Ion-driven gyroviscosity may dominate in the upper chromosphere and transition region. In strong fields ($\gtrsim 100\,  \mbox{G}$), viscosities primarily exceed diffusivities in the upper chromosphere and transition region. Parallel and perpendicular viscosities, being similar in magnitude, dampen waves and potentially compete with ambipolar diffusion in plasma heating, potentially inhibiting Hall and ambipolar instabilities when equal. The perpendicular viscosity tensor has two components,  $\nu_1$ and $\nu_2$, which differ slightly and show weak dependence on ion magnetization. Their differences, combined with shear, may destabilize waves, though magnetic diffusion introduces a cutoff for this instability. In configurations with a magnetic field $\bf{B}$ having vertical ($b_z=B_z/|\bf{B}|$) and azimuthal ($b_y=B_y/|\bf{B}|$) components, and a wavevector $\bf{k}$ with radial ($\kx=k_x/|\bf{k}|$) and vertical ($\kz=k_z/|\bf{k}|$) components, parallel viscosity and Hall diffusion can generate the viscous-Hall instability. Gyroviscosity further destabilizes waves in the upper regions. These findings indicate that the solar atmosphere may experience various viscous instabilities, revealing complex interactions between viscosity, magnetic fields, and plasma dynamics across different atmospheric regions.
\end{abstract}


\section{Introduction}
The dynamics of the partially ionized solar atmosphere, which govern the transport of mass, momentum, and energy across its stratified and magnetized layers, are notably complex. This complexity arises due to two primary factors. On one hand, the ionization state of the gas shifts from weakly ionized to partially ionized as we move from the photosphere through the chromosphere to the transition region between the chromosphere and the corona. On the other hand, the magnetic field strength varies dramatically, ranging from near zero to kG levels in both the quiet and active regions of the Sun \citep{SL00, D06}.
 
For instance, the photosphere, which comprises the visible ($380-750\,\mbox{nm}$) solar surface, extends vertically about $500$ to $550$ km. Here, the temperature decreases from around $6000\,K$ at the base to $4100\,K$ at the temperature minimum. In this layer, the plasma remains weakly ionized. Above the photosphere lies the chromosphere, named for its pinkish hue from the $H\alpha$ (Balmer-$\alpha$) emission line at $656.3\,\mbox{nm}$. The chromosphere, with a thickness of about $1-2\,\mbox{Mm}$, starts from the weakly ionized temperature minimum and transitions to a partially ionized region at its top, where temperatures reach around $\sim 2\times 10^4\,K$. Throughout the chromosphere, temperature, density, ionization degree, and magnetic field strength vary significantly.
  
A narrow transition region, only a few tens of kilometers thick, separates the chromosphere from the corona. In this region, the plasma temperature rises sharply to $10^6\,K$, while the density drops from approximately $10^{11}$ to $10^{8-9}\,\mbox{cm}^{-3}$. Consequently, the atmosphere transitions from partially ionized to almost fully ionized. With increasing temperature and decreasing density, the solar atmosphere also becomes optically thin in the transition region, though this may not always hold true, particularly during energetic and impulsive events like solar flares \citep{K19}.
  
Moreover, relatively cool ($T\sim 10^4\,\mbox{K}$), dense ($\sim 10^{10}-10^{11}\,\cc$), large-scale ($5-10^2$\, Mm), and partially ionized ($0.1-1$) structures are also observed as $\mbox{H}{\alpha}$-emitting plasma embedded within the hotter corona.
  
Our understanding of the solar magnetic field and its role in the transport of mass, energy, and angular momentum is still in its early stages. The magnetic field, generated in the Sun’s interior, is measured with the highest accuracy at the visible surface, the photosphere, and governs many of the physical processes in the solar atmosphere. Convective motions, both small- and large-scale, in the photospheric layers shuffle magnetic field lines, contributing to the buildup of magnetic energy in the corona (Parker 1987). While the small-scale dynamics of the photosphere are primarily driven by granular convection, the chromosphere is dominated by sound waves and magnetohydrodynamic (MHD) waves. For recent reviews on these processes, see \cite{B18, S21, S22, S24}.

In the partially ionized solar atmosphere, where the degree of ionization varies, the presence of neutrals leads to collisional momentum exchange with ions. This exchange can dampen waves and locally heat the plasma \citep{BR65, S22}. However, when the neutral-ion collision frequency is high compared to the signal frequency—i.e., when ions and neutrals are strongly coupled—ions acquire neutral inertia, leading to several important consequences \citep{PW06, PW08, P13, PW22} (hereafter PW22) :\\
1. A single-fluid, MHD-like description of the partially ionized plasma becomes valid in this regime.\\
2. The phase speed of the \alf wave is reduced compared to the fully ionized case.\\
3. The ion-cyclotron frequency is re-scaled due to ion-mass loading and becomes a much lower Hall frequency.\\
4. The ion-Larmor radius is re-scaled and becomes significantly larger than in a fully ionized plasma.

Although partially ionized solar plasma can be treated using an MHD-like, single-fluid framework, the magnetic field is not "frozen" in the plasma. It slips through due to collisions between electrons, ions, and neutrals, manifesting as Ohm, Hall, and ambipolar diffusion in the induction equation.

While Hall diffusion causes a dissipationless transport of the magnetic field, Ohm and ambipolar diffusion lead to energy dissipation. The role of ambipolar diffusion, in particular, has been a focus of recent research as a possible source of non-thermal heating in solar plasma \citep{D11, M11, Z11a, Z11b, KC12, Z12, Z13, L14, Ga14, S09, S15, CK15, S16, K17, M17, MS17, CK18, RC19, M21, K21, RC21, MS23, H24, M24}.

Magnetic field  mediates vortex motion in the photosphere, chromosphere and lower corona \citep{T23, B23}. 
Photospheric vortical motion can generate observable corotating structures in the chromosphere and corona, known as {\it chromospheric swirls} and {\it magnetic tornadoes}, spanning a wide range of spatial scales and extending from the upper convection zone to the transition region and lower corona \citep{KW17, K23}.  Vortices and flows of various spatial and temporal scales are observed in both active and quiet phases of the solar atmosphere \citep{B08, W09, Ba10, B10}. Bright points associated with vortex motion in the intergranular lanes typically move at speeds of $\lesssim 2\,\mbox{km}/\mbox{s}$ \citep{W09}.

The formation of small-scale intergranular vortices suggests that vorticity arises from the interaction between photospheric plasma and the ambient magnetic field in intergranular lanes \citep{M11, S11}. Non-ideal MHD simulations indicate that Hall diffusion generates out-of-plane velocity fields with maximum speeds of $\sim 0.1\,\mbox{km}/\mbox{s}$ at interface layers between weakly magnetized light bridges and neighboring strong-field umbral regions \citep{C12}. In summary, both observational evidence and numerical simulations highlight the presence of shear flows at various spatial scales within the solar photosphere.

The presence of large-scale shear flows can readily destabilize waves. For instance, the Kelvin-Helmholtz instability (KHI), which converts shear flow energy into vortex kinetic energy, has been proposed to explain the instability of flux tubes \citep{S10, Z10, K12}. Additionally, magnetic field diffusion \citep{PW12, PW13} and viscous momentum transport (PW22) can also drive non-ideal shear instabilities.

This work presents a detailed investigation of various viscous instabilities, building upon our previous study (PW22) but with several key differences: (i) Unlike PW22, which assumed a vertical magnetic field and transverse fluctuations (vertical wavevector), the current study considers a more general magnetic field topology and oblique wavevectors. Overall, this study offers a broader and more general framework compared to PW22.

It is worth noting that in this work, we describe the partially ionized solar plasma using an MHD-like framework. However, high-frequency, short-wavelength (on the order of a few meters) electrostatic waves can also be modeled using a multifluid approach \citep{G14}. Given the current observational resolution of approximately 90 km, only MHD waves are directly relevant to observations. Nevertheless, electrostatic fluctuations may still have an indirect effect on the MHD waves propagating through the medium.

The paper is organized as follows: Section 2 presents the basic set of equations and the dispersion relation. Subsection 2.1 describes the model solar atmosphere, followed by Subsection 2.2, which details the basic equations. In Subsection 2.3, the general dispersion relation is provided, along with an analysis of wave heating effects due to non-ideal MHD processes. Section 3 discusses the necessary conditions for the onset of viscous instabilities in the fluid and the role of magnetic diffusion in these instabilities. Section 4 addresses the application of the results, and Section 5 offers a brief summary of the findings.

\section{Basic set of equations and waves and instabilities in the medium}

\subsection{Model Atmosphere}
Collisions between ion and neutral particles also facilitate parallel, perpendicular, and cross viscous momentum transport relative to the magnetic field direction \citep{BR65, Z02}. As noted in PW22, the parallel and perpendicular neutral viscosities are of the same order, while the cross-neutral viscosity is negligible. The total viscosity in the solar atmosphere arises from both ions and neutrals. The primary contributors to the total viscosity are the parallel and perpendicular viscosities of the neutrals, along with the ion-induced gyroviscosity. As we will demonstrate, parallel and perpendicular viscosities play a significant role in the photosphere and chromosphere, whereas gyroviscosity becomes important in the upper chromosphere and transition region.

Given the highly diffusive nature of the solar atmosphere, it is essential to assess the relative importance of various viscosities compared to magnetic diffusivities. To quantify this, understanding the magnetic field distribution on the solar surface is crucial, where fields are organized into network and internetwork elements. The network magnetic field ($\gtrsim \mbox{kG}$) is primarily vertical and concentrated in flux tubes (diameter $\lesssim 100\,\mbox{km}$) located in intergranular lanes, whereas the internetwork field ($\sim \mbox{few}\,\mbox{G} - \mbox{kG}$), found in the interiors of supergranule cells, is predominantly horizontal \citep{H09, L08}.

In the lower chromosphere, strong ($\sim \mbox{kG}$) vertical flux tubes in network regions appear as bright points. These flux tubes, with a low filling factor (less than ($< 1 \%$) near their footpoints in the photosphere, expand to fill approximately $15 \%$ of the lower chromosphere (around $\sim 1 \mbox{Mm}$ in height, where CaII H and K emission lines are observed) before extending to fill the entire atmosphere as a canopy. The quiet solar internetwork region is also magnetized, with kG patches of field concentration and bright points similar to the network fields, though with an order of magnitude weaker fields elsewhere \citep{D09}. 

The variation of magnetic field strength with height can be inferred from pressure balance models in thin flux tubes, e.g., \citep{P79, G00, VK13, K15}.
\bq
B=B_0\,\exp{-\frac{z}{h}}\,,
\eq
where $B_0$ is the magnetic field at the footpoint, and $z/h$ is the height in the unit of pressure scale height, $h$. In the present work we adopt an alternative height variation \citep{M97} 
\bq
B = B_0\,\left(\frac{n_n}{n_0}\right)^{0.3}\,,
\label{eq:scl}
\eq 
where $n_n$ is the neutral number density, and $n_0$ is the reference number density at the footpoint. This relation ties the magnetic field variation to the density variation of neutral particles.

The magnetization of ions and electrons, quantified by the ion and electron Hall parameter $\beta_j$ is the ratio of the cyclotron frequency $\omega_{cj}$ to the plasma-neutral collision frequency $\nu_{jn}$. The cyclotron frequency is defined as:
\bq
\omega_{cj}=\frac{|q|\,B}{m_j\,c}\,,
\label{eq:cycf}
\eq
where $q$ is the charge, $B$ is the magnetic field strength, $m_j$ is the mass of the particle (either ion or electron), and 
$c$ is the speed of light. The Hall parameter is then expressed as:
\bq
\beta_j=\left(\frac{\omega_{cj}}{\nu_{jn}}\right)\,,
\label{eq:IhB}
\eq  
where $j=e\,,i$, referring to electrons and ions, respectively.

Parallel and perpendicular viscous momentum transport, which are of comparable magnitude, may compete with Ohmic and ambipolar diffusion of the magnetic field in the photosphere and chromosphere. In the upper chromosphere and the transition region between the chromosphere and corona, gyroviscous momentum transport becomes dominant over ambipolar diffusion. 

The Prandtl number, which is the ratio of viscosity to magnetic diffusivity, determines the relative importance of viscous transport over magnetic diffusion. Using the magnetic field profile, Eq.~(\ref{eq:scl}) we compare the Ohmic diffusivity ($\eta_O$), ambipolar diffusivity ($\eta_A$) and Hall diffusivity ($\eta_H$) with the parallel viscosity ($\nu_0$) and gyroviscosity ($\nu_3$). To compute these diffusivities and viscosities, we utilize the density and temperature data from \cite{F93} (hereafter F93), as was done in PW22. The definitions of $\eta$s and $\nu$s are provided in PW22 and listed in Table 1. We thus define the following Prandtl numbers  
\bq
Pr=\frac{max(\nu_0\,,\nu_3)}{max(\eta_O\,,\eta_H\,,\eta_A)}\,,
\label{eq:prn}
\eq
which compares the maximum of parallel ($\nu_0$) and gyro ($\nu_3$) viscosity with the maximum  of Ohm ($\etaO$), Hall ($\etaH$) and ambipolar ($\etaA$) diffusivities. 

\begin{figure}
\includegraphics[scale=0.35]{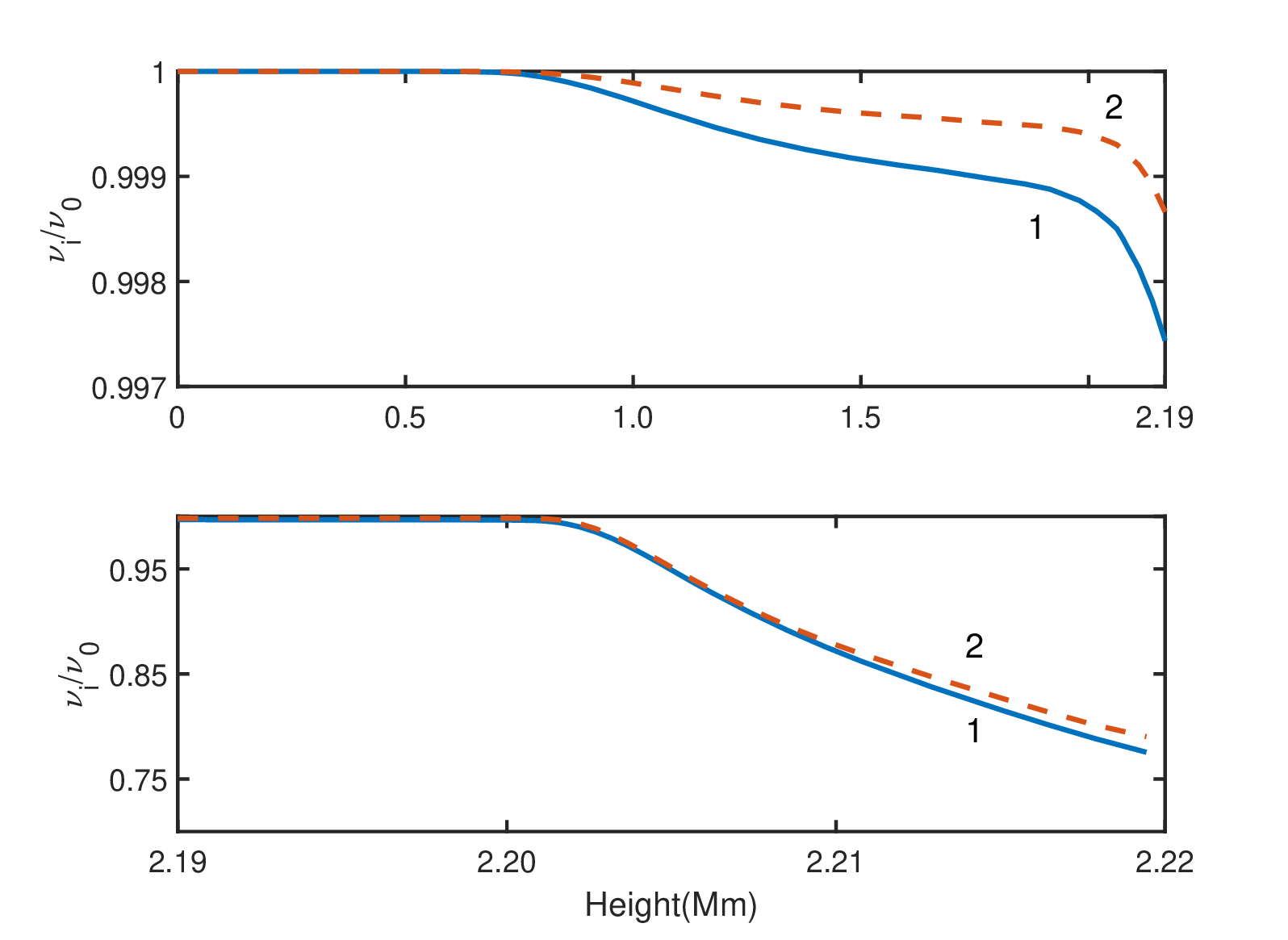}
\caption{The ratios $\nu_1/\nu_0$ (solid curve 1) and $\nu_2/\nu_0$ (dashed curve 2) are plotted as functions of height for the photosphere-chromosphere region (top panel) and the chromosphere-transition region (bottom panel). These ratios represent the relative magnitudes of different viscous transport coefficients at various altitudes. The altitude dependence of the magnetic field is derived from Eq.~(\ref{eq:scl}). The density and temperature profiles are taken from Table 2 (model C) of F93, which provides the relevant atmospheric data for these regions.}
\label{fig:F1}  
\end{figure}

Since the perpendicular viscosities, $\nu_1$, and $\nu_2$ are of the same order as the parallel viscosity $\nu_0$ and exhibit only a  weak dependence on ion-Hall $\beta_i$ [Fig.~(\ref{fig:F1})], the above Prandtl numbers also implies the relative importance of perpendicular viscosities against magnetic diffusivities.  Note that the slight difference between  $\nu_1$, and $\nu_2$  in Fig.~(\ref{fig:F1}) which results from their weak dependence on ion magnetization, becomes relevant in the chromosphere. As we will demonstrate below, this small difference is sufficient to destabilize waves in the presence of free shear energy.

\begin{figure}
\includegraphics[scale=0.35]{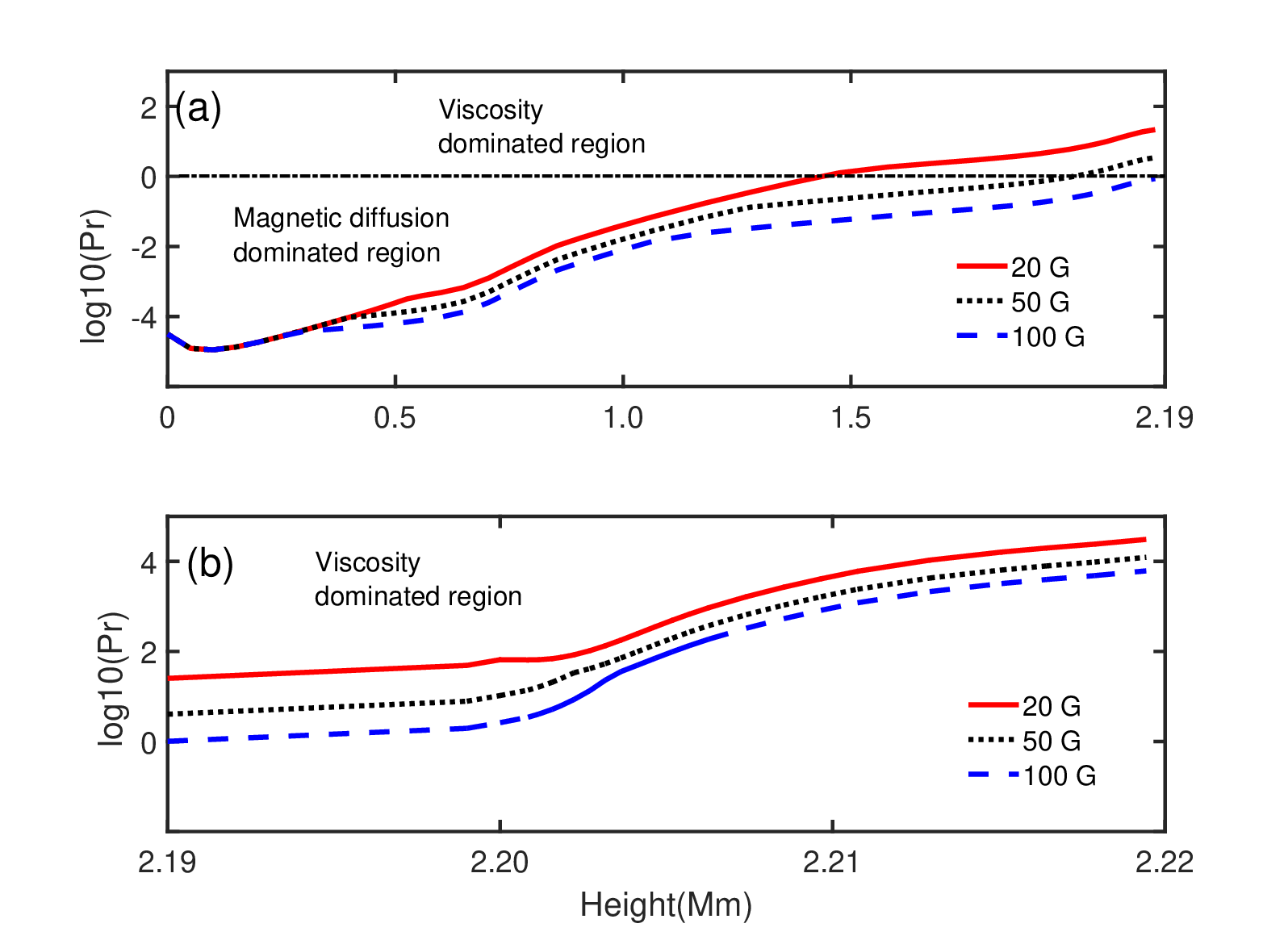}
\caption{The Prandtl number is plotted as a function of height for three different magnetic field strengths: 
$B_0=20\,,50\,100\,\mbox{G}$. These values correspond to varying magnetic field strengths at the footpoint. The other parameters,  are the same as those used in the previous figure. 
}
 \label{fig:F2}  
\end{figure}

\begin{figure}
\includegraphics[scale=0.35]{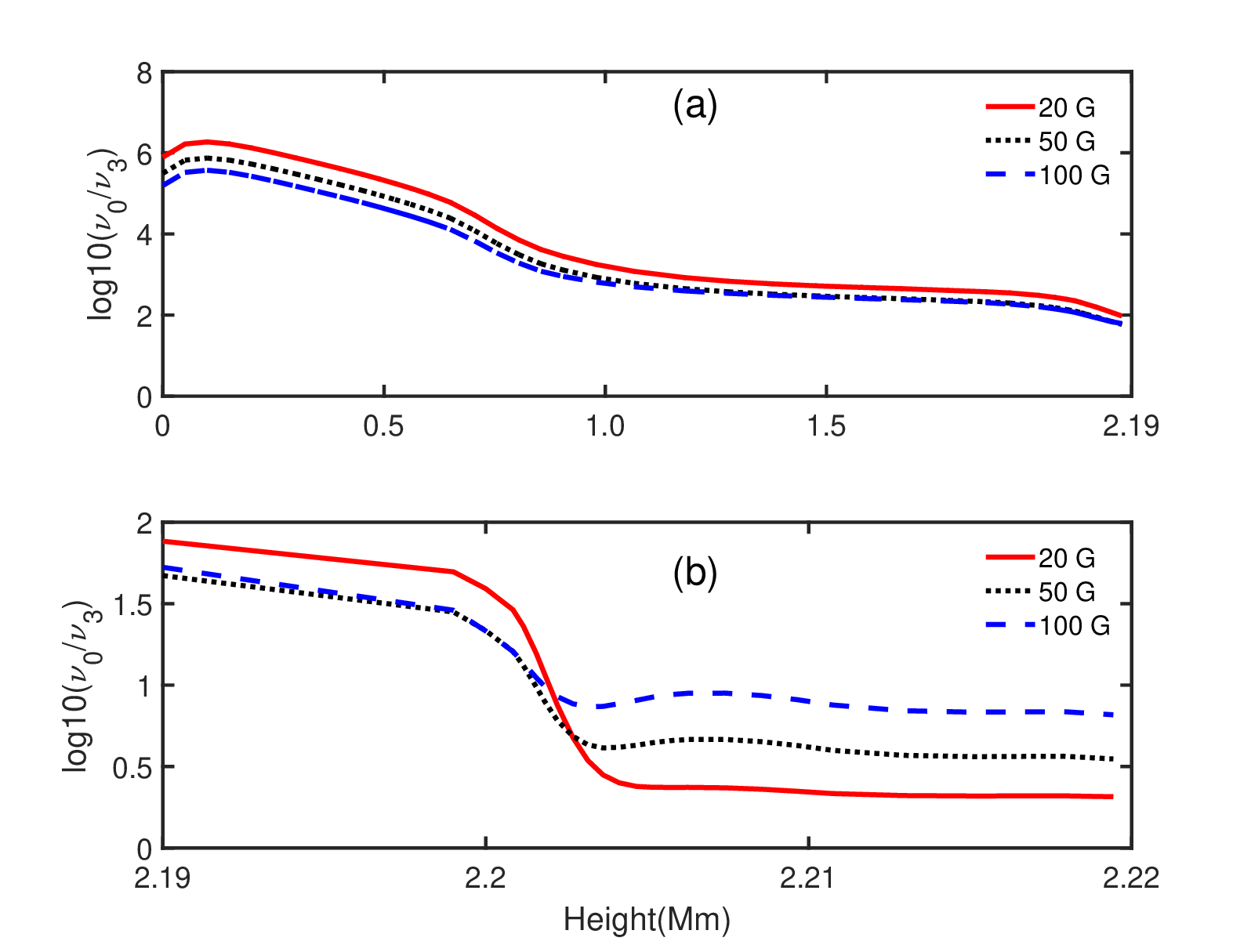}
\caption{Ratio of the parallel and gyro viscosity are plotted against height for $B_0=20\,,50\,,100\,\mbox{G}$. Other parameters are the same as used in the previous figure.}
 \label{fig:F3}  
\end{figure}

In Fig.~(\ref{fig:F2})  we plot Prandtl number, $Pr$ for $B_0=20\,,50\,,100\,\mbox{G}$. For $B_0=20\,\mbox{G}$ field, $Pr>1$ above an altitude of $1.4\,\mbox{Mm}$ [top panel (a), solid line]. This indicates that, for a weak magnetic field, viscous momentum transport dominates magnetic diffusion in the middle and upper chromosphere, as well as the transition region [lower panel (b), solid line]. Since parallel viscosity dominates over gyroviscosity [Fig.~\ref{fig:F3} (a) and (b), solid line], both parallel and perpendicular viscous momentum transport become the dominant non-ideal MHD effects in these regions. 

For a moderate magnetic field of $B_0=50\,\mbox{G}$ [dotted line in Figs.~(\ref{fig:F2})-(\ref{fig:F3})], $Pr>1$  is only observed near the upper chromosphere, around $\sim 2.18\,\mbox{Mm}$ and beyond. Thus, in the presence of a moderate-strength field, magnetic diffusion dominates over viscous diffusion in the photosphere and the lower and middle chromosphere, while viscous momentum transport becomes dominant in the upper chromosphere and transition region.

When $B=100\,\mbox{G}$ [dashed line in Figs.~(\ref{fig:F2})-(\ref{fig:F3})], viscous momentum transport becomes significant only in the upper chromosphere and transition region. In this case, gyroviscosity plays an important role [Fig.~\ref{fig:F3}(b)].

Thus, the dynamics of the chromosphere can be categorized into two regimes:

For a weak magnetic field (e.g., in the quiet solar region), magnetic diffusion dominates viscous transport in the photosphere and middle chromosphere, while viscous transport overtakes ambipolar diffusion in the upper chromosphere and transition region [Fig.~\ref{fig:F2}(a)].

For a strong magnetic field ($\sim 100\,\mbox{G}-\mbox{kG}$, typical of active solar regions), magnetic diffusion dominates viscous transport throughout the entire photosphere-chromosphere. In the transition region [Fig.~\ref{fig:F2}(b)], where ambipolar diffusion is negligible, viscous transport becomes the only significant non-ideal mechanism.

In summary, viscous transport dominates over magnetic diffusion in the photosphere-chromosphere and coronal filaments only when the magnetic field is weak. As the magnetic field strength increases, the region where viscous momentum transport is important shifts to the upper chromosphere and transition region. In the presence of a strong field, viscous and gyroviscous effects are significant only in the transition region.

\begin{figure}
\includegraphics[scale=0.35]{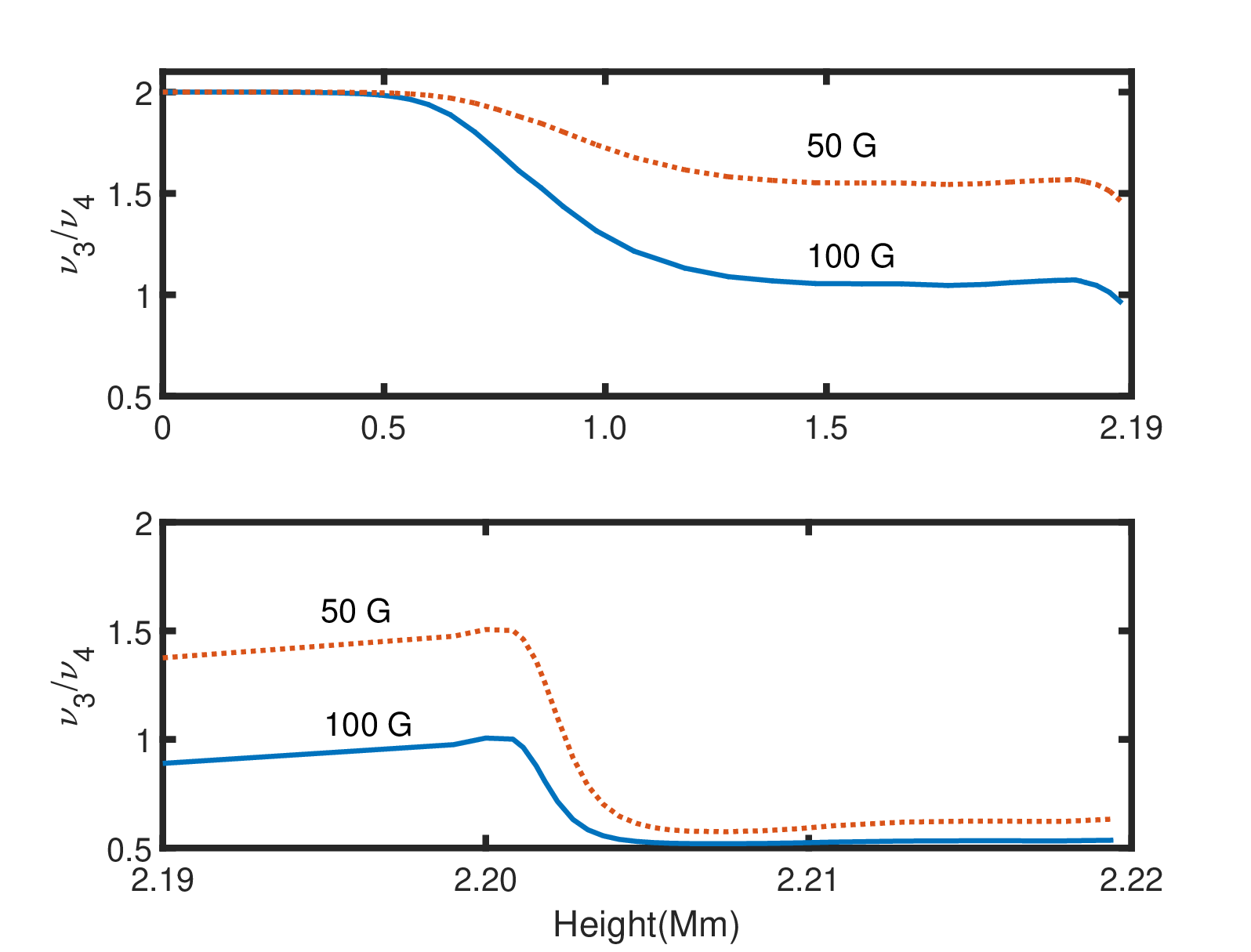}
\caption{The ratio of the gyro viscosities $\nu_3/\nu_4$ is plotted against height for $B_0=50$ and $100\,\mbox{G}$ field. Other parameters are the same as used in the previous figure.}
 \label{fig:F4}  
\end{figure}
The gyroviscosities, $\nu_3$ and $\nu_4$  depend on the ion-Hall parameter, $\beta_i$  and and vary throughout the solar atmosphere. Their ratio is given by:
\bq
 \frac{\nu_3}{\nu_4}=\frac{1}{2}\left(1+\frac{3}{1+\beta_i^2}\right)\,,   
\eq
and is shown in Fig.~(\ref{fig:F4}) for the magnetic field profile Eq.~(\ref{eq:scl}). 

Since gyroviscosity manifests only when ions are magnetized, it becomes important in the upper chromosphere ($\gtrsim 2.19,\mbox{Mm}$) and beyond, where $\beta_i>1$ is satisfied. In the transition region (lower panel of Fig.~(\ref{fig:F4})), the ratio $\nu_3/\nu_4$ asymptotically approaches $1/2$ for both $B_0=50$ and $100\,\mbox{G}$ fields, as $\beta_i\gg1$ in this region. In the upper chromosphere, at the base of the transition region, $\nu_3\approx \nu_4$ when $B_0=100\,\mbox{G}$, while $\nu_3\approx 1.5\, \nu_4$ when $B_0=50\,\mbox{G}$.
\begin{table*}
\centering
\caption{List of Frequently Used Symbols}
\label{tab:symbols}
\begin{tabular}{llll}
\hline
Symbol & Definition & Symbol & Definition \\
\hline
$\kx$ & $k_x / |\k|$ & $\kz$ & $k_z / |\k|$ \\
$b_y$ & $B_y /|\B|$ & $b_z$ & $B_z/|\B|$ \\
$\bb$ & $\B/|\B|$ & $g$ & $-\kx\,\kz\,b_y\,b_z$ \\
$\kh$ & $\k/ |\k|$ & $\mu$ & $\kh\cdot \bb$ \\
$\omega_{cj}$ & Cyclotron frequency (j=ion, electron) & $\nu_{jn}$ & Collision frequency \\
$\nu_n$ & Total neutral collision frequency & $\Gamma_P\,,\Gamma_{vis}$ & Damping rates \\
$\beta_j$ & Electron and ion-Hall parameter & $Pr$ & Prandtl number \\
$Pr_{\eta_j}$ & \multicolumn{3}{l}{$(\nu_0+\nu_1+\nu_2)/\eta_j$ (j=Ohm, ambipolar, Hall, Pedersen Prandtl number)} \\
$\va$ & Alfvén speed & $\omega_A$ & $k\,\va$ (Alfvén frequency) \\
$c_s$ & $\sqrt{\frac{k_B\,T}{m_i}}$ (sound speed) & $\beta$ & $\sqrt{\frac{2\,c_s^2}{\va^2}}$ (plasma beta) \\
$\etaO$ & Ohm diffusivity & $\etaA$ & Ambipolar diffusivity \\
$\etaP$ & $\etaO + \etaA$ (Pedersen diffusivity) & $\etaH$ & Hall diffusivity \\
$\nu_0$ & Parallel viscosity & $\nu_{1\,,2}$ & Perpendicular viscosity \\
$\nu_3\,,\nu_4$ & Gyro viscosity & $\alpha$ & $\nu_3/\nu_4$ \\
\hline
\end{tabular}
\end{table*}
\subsection{Basic set of equations}
The plasma in the photosphere, chromosphere, and transition region consists of electrons, protons, singly ionized metallic ions, neutral hydrogen (H), helium in its neutral (He I), singly ionized (He II), and doubly ionized (He III) states. For simplicity, and ignoring the distinction between hydrogen and metallic ions, we assume that the partially ionized plasma is composed of electrons, singly charged ions, and neutral hydrogen.

The dynamics of this partially ionized solar plasma in the photosphere-chromosphere region can be described by a modified set of magnetohydrodynamic (MHD) equations (PW22). These equations take into account the interaction between ionized and neutral components, as well as non-ideal MHD processes such as resistivity, Hall effects, and ambipolar diffusion. The governing equations are as follows:
\bq
\frac{\partial \rho}{\partial t} + \grad\cdot\left(\rho\,\v\right) = 0\,.
\label{eq:cont}
\eq
Here $\rho = \rho_i + \rho_n$ is the bulk mass density and  $\rho_{i\,, n} = m_{i\,,n}\,n_{i\,,n}$ is the ion and neutral mass densities with $m_{i\,,n}\,, n_{i\,,n}$ as the ion and neutral mass and number densities respectively; $\v = (\rho_i\,\vi + \rho_n\,\vn)/\rho$ is the bulk velocity, and, $\vi$ and $\vn$ are bulk velocities of the ion and neutral fluids respectively.
\bq
\rho\,\frac{d\v}{dt}=  - \nabla\,P - \nabla\cdot \tenq{\Pi} + \frac{\J\cross\B}{c}\,,
\label{eq:meq}
\eq
where $\J = e\,n_e\,\left(\vi - \ve\right)$ is the current density, $\B$ is the magnetic field and $P = P_e + P_i + P_n$ is the total pressure and the non--diagonal viscous stress tensor is \citep{BR65, C86}  
\bq
\tenq{\Pi} = \tenq{\Pi}_{\parallel} +  \tenq{\Pi}_{\perp} + \tenq{\Pi}_{\Lambda}\,,
\eq
where $\parallel\,,\perp\,,\Lambda$ are the parallel [$\bb(\bb\cdot\nabla)$], perpendicular [$-\bf{b}\cross(\bf{b}\cross \nabla$)] and cross ($\bf{b}\cross \nabla$) terms with respect to the magnetic field direction $\bb=\B /B$. The above stress tensor are related to the strain tensor
\bq
\tenq{\Pi} = - \rho\nu_0\,\tenq{W}_{0} - \rho\nu_1\,\tenq{W}_{1} - \rho\,\nu_2\,\tenq{W}_{2} + \rho\nu_3\,\tenq{W}_{3} + \rho\,\nu_4\,\tenq{W}_{4}\,,
\label{eq:vst}
\eq
where $\nu_0$ represents the parallel viscosity, $\nu_1$, and $\nu_2$ are the perpendicular viscosities and $\nu_3$ and $\nu_4$ denote the gyroviscosities. Expressions for the ion and neutral viscosity coefficients when $T_i = T_e= T_n = T$ are given in \cite{PW22}.
Note that 
\bq
\nu_0 \approx \nu_1 \approx \nu_2\,,\nu_3 \approx \nu_0/\beta_i\,.
\eq
Further $\tenq{W}_{0}\,,\tenq{W}_{1}\,,\tenq{W}_{2}\,,\tenq{W}_{3}$, and $\tenq{W}_{4}$ are related to stress $\tenq{W}$
\bq
\tenq{W} = \grad\v + \left(\grad\v\right)^{T} - \frac{2}{3}\,\tenq{I}\div\v\,,
\label{eq:stn}
\eq
via 
\begin{eqnarray}
\tenq{W}_{0}&=&  \frac{3}{2}\,\left(\bb\cdot\tenq{W}\cdot\bb\right)\left(\bb\bb-\frac{1}{3}\tenq{\bf{I}}\right)\,,
\nonumber\\
\tenq{W}_{1}&=&\tenq{\bf{I}}_{\perp}\cdot\tenq{W}\cdot \tenq{\bf{I}}_{\perp}
-\frac{1}{2}\tenq{\bf{I}}_{\perp} \tenq{\bf{I}}_{\perp}\bmath{:}\tenq{W}\,,
\nonumber\\
\tenq{W}_{2}&=&\tenq{\bf{I}}_{\perp}\cdot\tenq{W}\cdot \bb\bb + \bb\bb \cdot \tenq{W}\cdot  \tenq{\bf{I}}_{\perp}\,,
\nonumber\\
\tenq{W}_{3}&=&\frac{1}{2}\left(\bb\cross\tenq{W}\cdot\tenq{\bf{I}}_{\perp} - \tenq{\bf{I}}_{\perp}\cdot \tenq{W}\cross\bb \right)\,,
\nonumber\\
\tenq{W}_{4}&=& \bb\cross\tenq{W}\cdot\bb\bb - \bb\bb\cdot\tenq{W}\cross \bb\,.
\label{eq:st1}
\end{eqnarray}
Here, the projection tensor $\tenq{\bf{I}}_{\perp}$ defines the projection in the plane perpendicular to the magnetic field and is given by $\tenq{\bf{I}}_{\perp}=\tenq{\bf{I}}-\bb\bb$. Here $\tenq{\bf{I}}$ is the full identity tensor  and $\bb\bb$ is the dyadic product (outer product) of the unit vector $\bb$  with itself, which isolates the component along the magnetic field. Thus,  $\tenq{\bf{I}}_{\perp}$ effectively removes the component along the magnetic field direction, leaving only the perpendicular components. This tensor is crucial when describing processes like perpendicular viscosity, where the transport of momentum is constrained by the magnetic field. 

Defining 
\bq
\alpha=\frac{\nu_3}{\nu_4}\,,
\label{eq:alp}
\eq 
the gyroviscous terms $\nu_3\,\tenq{W}_{3}+\nu_4\,\tenq{W}_{4}$ can be combined together as 
\bq
\nu_3\,\left(\tenq{W}_{3}+\frac{1}{\alpha}\,\tenq{W}_{4}\right)=\frac{\nu_3}{2}\,\left(\bb\cross\tenq{W}\cdot\left(\tenq{\bf{I}}+N\,\bb\bb\right) + \left(\right)^{\mbox{T}}\right)\,. 
\eq
Here $\left(\right)^{\mbox{T}}=-\left(\tenq{\bf{I}}+N\,\bb\bb\right)\cdot\tenq{W}\cross\bb$ is the transpose of the first term and
\bq
N=\frac{2}{\alpha}-1\,.
\eq
In the component form $\tenq{W}$ is
\begin{eqnarray}
W_{xx}&=&\frac{4}{3}\partial_x\vx-\frac{2}{3}\left(\partial_y\vy+\partial_z\vz\right)\,, 
\nonumber\\
W_{yy}&=&\frac{4}{3}\partial_y\vy-\frac{2}{3}\left(\partial_x\vx+\partial_z\vz\right)\,,
\nonumber\\
W_{zz}&=&\frac{4}{3}\partial_z\vz-\frac{2}{3}\left(\partial_x\vx+\partial_y\vy\right)\,, 
\end{eqnarray}
and the remaining symmetric part is 
\begin{eqnarray}
W_{xy} &=& \partial_x\vy+\partial_y\vx\,, 
\nonumber\\
W_{xz} &=& \partial_x\vz+\partial_z\vx\,, 
\nonumber\\
W_{yz}&=& \partial_y\vz+\partial_z\vy\,. 
\end{eqnarray}

The induction equation is \citep{PW08}
\begin{eqnarray}
\delt \B = \curl\left[
\left(\v\cross\B\right) - \frac{4\,\pi\,\eta_O}{c}\,\J - \frac{4\,\pi\,\eta_H}{c}\,\J\cross\bb
\right. \nonumber\\
\left.
+ \frac{4\,\pi\eta_A}{c}\,
\left(\J\cross\bb\right)\cross\bb
\right]\,,
\label{eq:ind}
\end{eqnarray}
Eqs.~(\ref{eq:cont}), (\ref{eq:meq}) and (\ref{eq:ind}) together with the barotropic relation $P = c_s^2\,\rho$ and Amp\'ere's law 
\begin{equation}
    \J = \frac{c}{4\pi}\curl\B\,.
    \label{eq:Amp}
\end{equation}
completes the single fluid MHD description with the FLR correction of a partially ionized plasma. The ambipolar and Hall diffusion coefficients can be expressed in terms of the Ohm diffusion coefficient via the electron Hall parameter, $\beta_e$, and ion-Hall parameter $\beta_i$ as:
\bq
\eta_A= \beta_e\,\beta_i\,\eta_O\,,\quad \eta_H=\beta_e\,\eta_O\,.
\eq
Thus, ambipolar diffusion dominates over Hall when $\beta_i>1$, indicating that ions are sufficiently magnetized. In this regime, the Lorentz force strongly influences ion dynamics, making ambipolar diffusion-which arises from the decoupling between neutral and ionized species-the dominant non-ideal MHD process.

Both gyroviscous momentum transport and ambipolar diffusion operate in the regime where $\beta_i>1$. When ions are magnetized (i.e., when the ion cyclotron frequency exceeds the ion-neutral collision frequency), both effects become relevant. Gyroviscosity becomes important because the ion motion perpendicular to the magnetic field experiences significant deflection, while ambipolar diffusion becomes crucial due to the 
differential motion between the ionized and neutral components.

\subsection{Waves in the solar atmosphere}
Various wave modes including fast kink and torsional \alf mode have been observed in the long-lived vortex flows \citep{T20}. The detection of transverse waves in umbral fibrils in the chromosphere of a strongly magnetized ($\sim 3-5\,\mbox{kG}$) sunspot \citep{Y23} suggests that these waves have energy flux $\sim 7.52\times 10^{11}\,\mbox{erg}\,\mbox{cm}^{-2}$. This energy flux is three to four orders of magnitude higher than the radiative losses of the coronal plasma, which are estimated at around $10^{8}\,\mbox{erg}\,\mbox{cm}^{-2}$. This indicates that such waves could play a significant role in energy transport, contributing to the heating of the solar corona beyond what radiative processes alone can account for.

When is the viscosity important for wave damping in the plasma? To understand this, we take the ratio of the inertial term to the viscous stress tensor $\tenq{\Pi}$ in the momentum equation (\ref{eq:meq}), and define the Reynolds number as:
\bq
R_1=\frac{\rho\,\omega\,V\,L}{|\tenq{\Pi}|} \sim \frac{\omega}{k\,c_s}\frac{\nu_n}{k\,c_s}\,,
\label{eq:FRN}
\eq
where $\rho$ is the plasma density, $\omega$ is the wave frequency, $V$ is the characteristic velocity, $L$ is the characteristic length, and $k$ is the wavenumber. Here we have used $L\sim 1/k$ and $V\sim \omega/k$ with wave frequency $\omega$ and wave number $k$. The sound speed $c_s$ serves as a characteristic velocity scale for the medium. Further, $\nu_n$ is the total neutral collision frequency, which is given by the following expression (PW22)
\bq
 \nu_n = 0.3\,\nu_{nn} + 0.36\,\nu_{ni} + 0.4\times 10^{-3}\,\nu_{ne}\,,   
\eq
where $\nu_{nn}$, $\nu_{ni}$, and $\nu_{ne}$ represent neutral-neutral, neutral-ion, and neutral-electron collision frequencies, respectively. 

When $R_1$ is small, viscosity dominates, and the damping of the wave is significant. Conversely, when $R_1$ is large, the inertial effects dominate, and the viscous damping is less relevant. Thus, viscosity strongly influences wave behavior, particularly in the chromosphere and transition regions, where collisions between neutrals and ions are frequent and the Reynolds number can drop, leading to more effective viscous wave damping.

In the gyroviscous case, we compare the inertia and the gyroviscous stress term in the momentum equation  (\ref{eq:meq}) and obtain the following Reynolds number \citep{Y66}:
\bq
R_2=\frac{\rho\,\omega\,V\,L}{|\tenq{\Pi}_{\Lambda}|}\sim \frac{\omega}{\omega_H}\left(k\,R_L^*\right)^{-2} \sim \frac{\omega}{\omega_H}\,\beta_i^2 \,,
\eq
where $\omega_H\,,R_L^*$ are the Hall frequency and Larmor radius, respectively (PW22). Since $\beta_i>1$ in the chromosphere and transition region, $R_2\sim \mathcal{O}(1)$ only for waves with frequencies $\omega<\omega_H$.  In these regions $\omega_H$ approaches the ion cyclotron frequency $\omega_{ci}$, especially in the upper chromosphere and transition region, where the neutral population is depleting. Because $\omega<\omega_H$ is easily satisfied under these conditions, this implies that gyroviscosity plays a significant role in the upper chromosphere and transition region of the solar atmosphere. 
  
As the focus of present investigation is low frequency behaviour of the medium, we shall work in the  Boussinesq approximation, i.e. assume that sound waves propagate fast enough so that the fluid is nearly incompressible. Consequently, the phase speed of the waves is much smaller than the sound speed, $c_s$ or $\omega\ll k\,c_s$. Given that $k\,c_s\ll \nu_n$, it follows from Eq.~(\ref{eq:FRN}) that both parallel and perpendicular viscosity corrections to the momentum equation become significant in this approximation. 

After linearising  the continuity, momentum and induction equations and  assuming an axisymmetric perturbations of the form $\exp \left(i \,\k  \cdot {\bf{x}} + \sigma \, t\right)$, 
 where $\sigma=-i\,\omega$ and $k = (k_x , 0 , k_z)$, we get the following dispersion relation (appendix B) 
\bq
\sigma^4+(C_3+E_3)\sigma^3+(C_2+E_2)\sigma^2+(C_1+E_1)\sigma+(C_0+E_0)=0\,,
\label{eq:mE}
\eq
where $C_j$ contains only viscous terms, while $E_j$ contains diffusion and mixed terms [Eqs.~(\ref{eq:CoE1}) and (\ref{eq:deE}) in the appendix B].

\subsection{Wave propagation in the absence of shear flow, i.e. $s=0$}
Without shear flow, solar atmospheric waves are damped by viscosities and magnetic (Ohm and ambipolar) diffusivities. Setting the shear gradient $s=0$ in the dispersion relation, we analyze wave dissipation into heat, including chromospheric heating via ambipolar diffusion. 
Following [section 8\,, \cite{BR65}], the rate of dissipation of wave energy equated to plasma heating yields:
\bq
2\,\Gamma\,\rho\,\va^2\,\Bigg[
\left(\frac{\overline{\delta\,B}}{B}\right)^2+\frac{\beta}{2}\left(\frac{\overline{\delta\,\rho}}{\rho}\right)^2
\Bigg]=\Theta\,T\,.
\label{eq:HtR}
\eq
Here $\Theta$ is the entropy production of the wave, T is the temperature and $\beta=2\,c_s^2/\va^2$ is the plasma beta parameter-a ratio of  thermal to magnetic energy densities. The bar over the quantities indicates volume averaging. While Eq.~(\ref{eq:HtR}) was originally derived for fully ionized plasmas, its form remains applicable to partially ionized plasmas. The heating mechanism described by this equation provides a useful estimate of how wave energy is converted into thermal energy in various plasma environments, including the partially ionized regions of the solar atmosphere. 

For the magnetosonic waves, magnetic and density fluctuations are related, $\delta\,\rho/\rho \sim \delta\,B/B$ and thus,
\bq
\Bigg[
\left(\frac{\overline{\delta\,B}}{B}\right)^2+\frac{\beta}{2}\left(\frac{\overline{\delta\,\rho}}{\rho}\right)^2
\Bigg] \sim \left(1+\frac{\beta}{2}\right)\,\left(\frac{\overline{\delta\,B}}{B}\right)^2\,.
\eq
In the chromosphere $1+\beta/2\sim 1$ and thus, heating solely depends on the magnetic power spectrum $\sim \delta\,B^2$.

As $\nu_0 \approx \nu_1 \approx \nu_2$,  the viscous damping rate  for the \alf ($b_y=0$) wave is \citep{BR65}
\bq
\Gamma_{\mbox{vis}} = k_{\perp}^2\,\nu_1 + k_{\parallel}^2\,\nu_2 =  k^2\,\nu_1\,.
\label{eq:vis1}
\eq
Damping of the magnetosonic ($b_y=0\,,\delta v_y=0$) wave is
\citep{BR65} 
\bq
\Gamma_{\mbox{vis}} = 
\left( 0.33\,\nu_0 + \nu_1\right)\,k_{\perp}^2 + k_{\parallel}^2\,\nu_2\approx \left(0.33\,\sin\theta + 1\right)\,k^2\,\nu_0\,.
\label{eq:mvis}
\eq
Here $k_{\perp}=k\sin\theta$. Evidently, the damping rate of the magnetosonic wave exceeds that of the \alf wave.

Comparing the viscous damping rate with the Pedersen ($\eta_P=\eta_O+\eta_A$) damping rate
\bq
\Gamma_{\mbox{P}}=k^2\,\eta_P\,,
\label{eq:pH}
\eq
we find that the Prandtl number $Pr_{\etaP}=(\nu_0+\nu_1+\nu_2)/\eta_P$ serves as a diagnostic to determine which process—viscous or resistive—dominates the heating of the partially ionized plasma. The total damping rate is the sum of both the Pedersen and viscous damping rates:
\bq
\Gamma\approx \left(1+ Pr_{\etaP}\right)k^2 \,\eta_P\,. 
\label{eq:ht}
\eq
\begin{figure}
\includegraphics[scale=0.34]{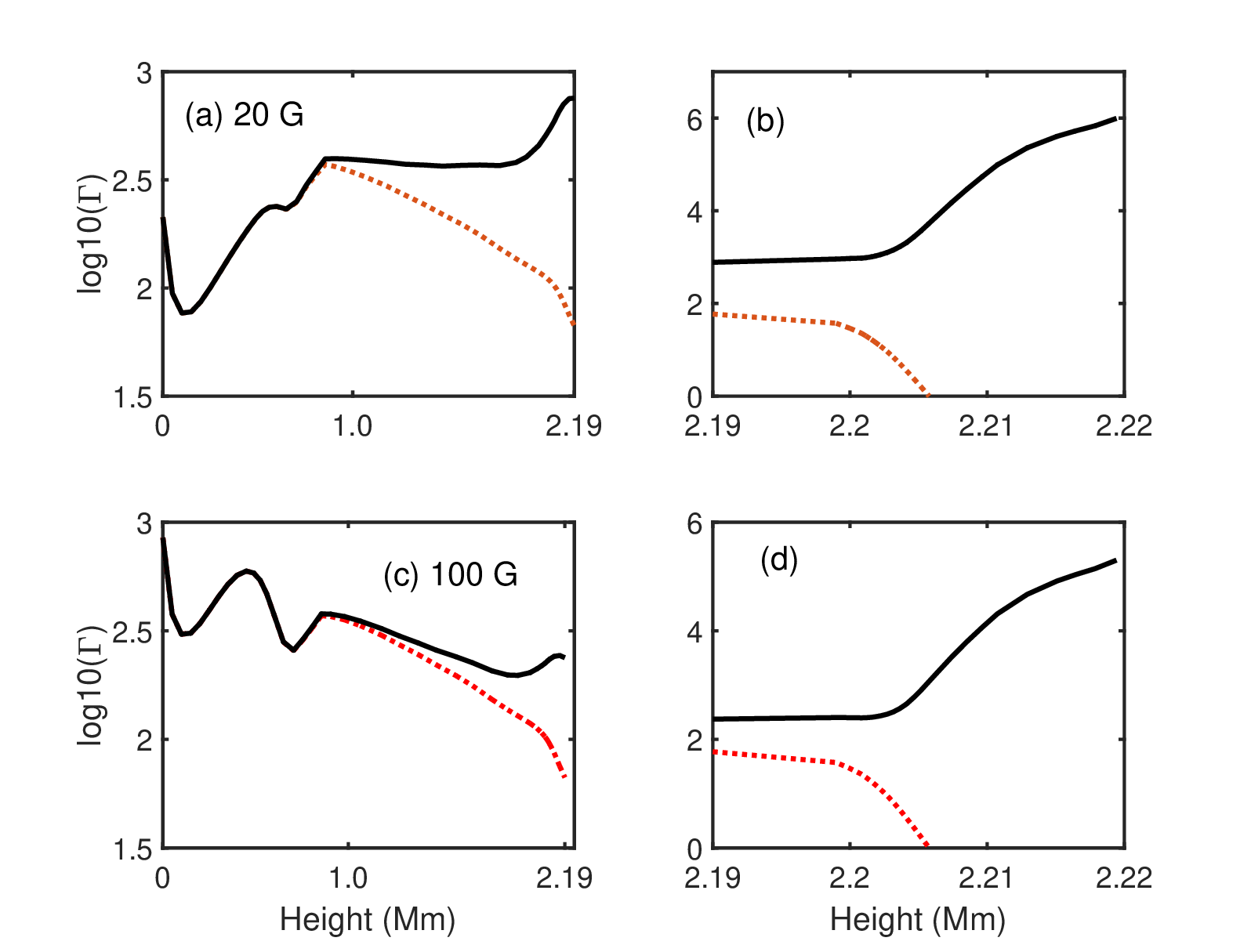}
\caption{The total damping rate $log10(\Gamma)$ [Eq.~(\ref{eq:ht})] (solid curve) and Pedersen damping rate  [Eq.~(\ref{eq:pH})] ) dotted curve) are plotted against height for $B=20$ G (top panel), and $100$ G (bottom panel).}
\label{fig:F5}  
\end{figure}

In Fig.~(\ref{fig:F5}) we plot the total damping rate (upper solid curves), $\Gamma$ and the Pedersen damping rate, $\Gamma_{\mbox{P}}$ (lower dotted curve) against height for $B_0=20$ G [Panels (a) and (b)], and $100$ G [panels (c) and (d)]. The value of the wavenumber is determined by the Pedersen cutoff,
\bq
k_P=k_A+k_O\,,
\eq
where the ambipolar and Ohm cutoff wavenumbers, 
\bq
k_A=\frac{\nu_{ni}}{\va}\equiv \frac{1}{L_A}\,,\quad k_O=\frac{\beta_e\,\beta_i}{L_A}
\eq
are expressed in terms of ambipolar length scale $L_A$. We see from Fig.~(\ref{fig:F5}) that in the case of a weak magnetic field ($B=20\,\mbox{G}$), which is characteristic of the quiet regions of the Sun, the total damping rate ($log10 (\Gamma)$) [solid curve in the panels (a) and (b)] is significantly larger than the ambipolar damping rate (dotted curves) in the upper chromosphere and transition region. This indicates that viscous damping, rather than ambipolar diffusion, plays a dominant role in heating these regions. However, for a strong magnetic field ($B=100\,\mbox{G}$), the total damping rate [solid curves in the panels (c) and (d)] is still larger than the ambipolar damping rate but not by orders of magnitude in the upper chromosphere. It is only in the transition region [panel (d)] where the total damping rate, becomes orders of magnitude higher than the ambipolar rate. This can be attributed to the sharp decline in neutral number density in the transition region, which significantly reduces ambipolar diffusion. Comparing the weak and strong field cases, we note that the viscous heating is more efficient in the presence of a weak field. This is because ambipolar damping rate varies as $\propto B^2$. 

\begin{figure}
\includegraphics[scale=0.34]{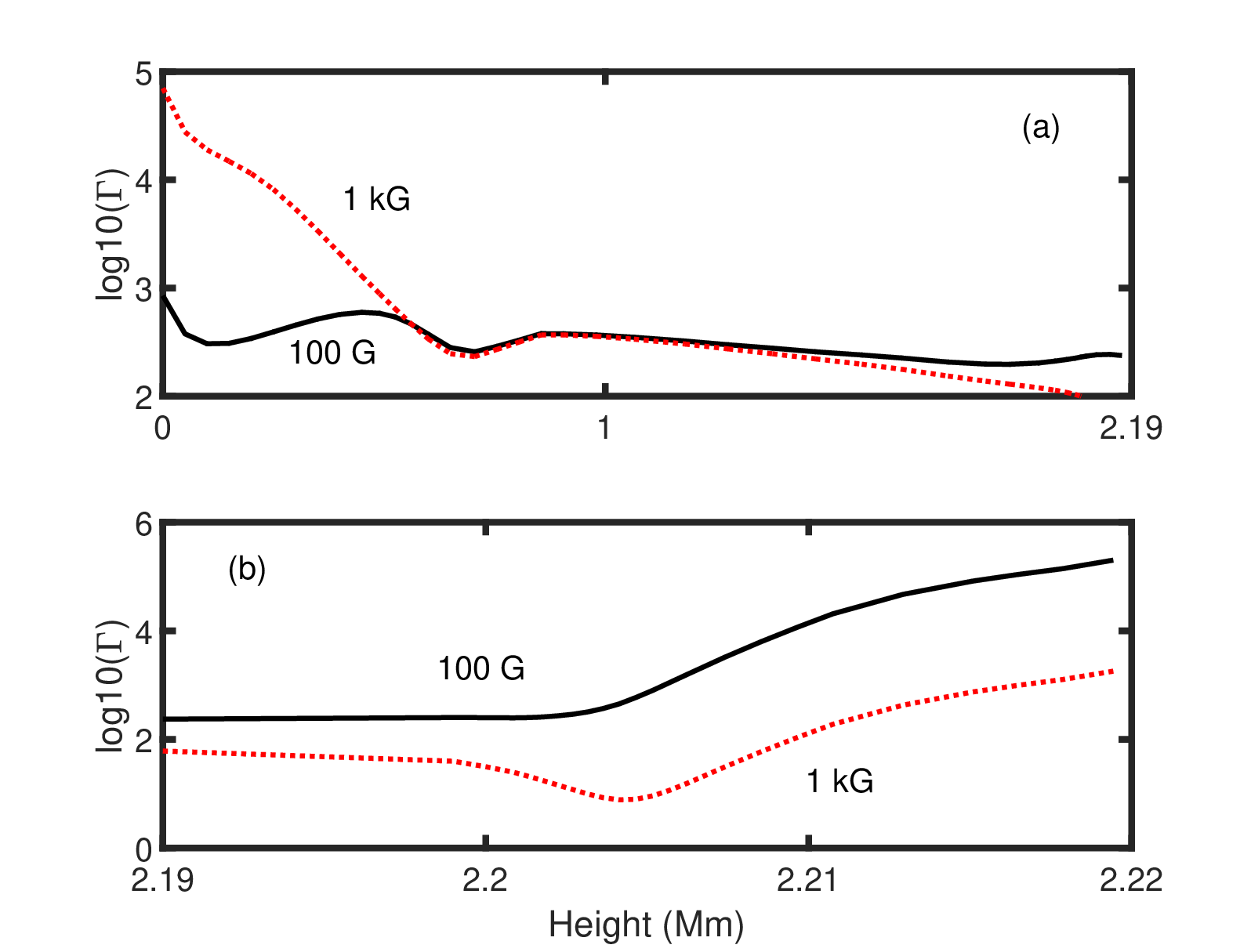}
\caption{The total damping rate $log10(\Gamma)$ for $1\,kG$ (dotted curve) and $100\,G$ field (solid curve).}
\label{fig:F6}  
\end{figure}

When the field is strong ($\gtrsim 1\,\text{kG}$), ambipolar and viscous heating rates follow the same dotted curve in Fig.\ref{fig:F6}, indicating equal heating efficiency in the middle chromosphere. In the upper chromosphere and transition region, Fig.\ref{fig:F6}(b) shows higher viscous heating for weaker field ($100~\text{G}$). The ambipolar heating rate remains constant for both $100$ and $\text{1\,kG}$ fields [Fig.~\ref{fig:F5}(d))]. In summary, solar atmospheric heating depends on magnetic field strength: in quiet regions ($\lesssim 100~\text{G}$), viscous damping dominates chromospheric heating, while strong field regions ($\gtrsim \text{kG}$) exhibit both mechanisms. In the upper chromosphere and transition region, where neutral density decreases, viscous heating prevails.

\begin{figure}
\includegraphics[scale=0.34]{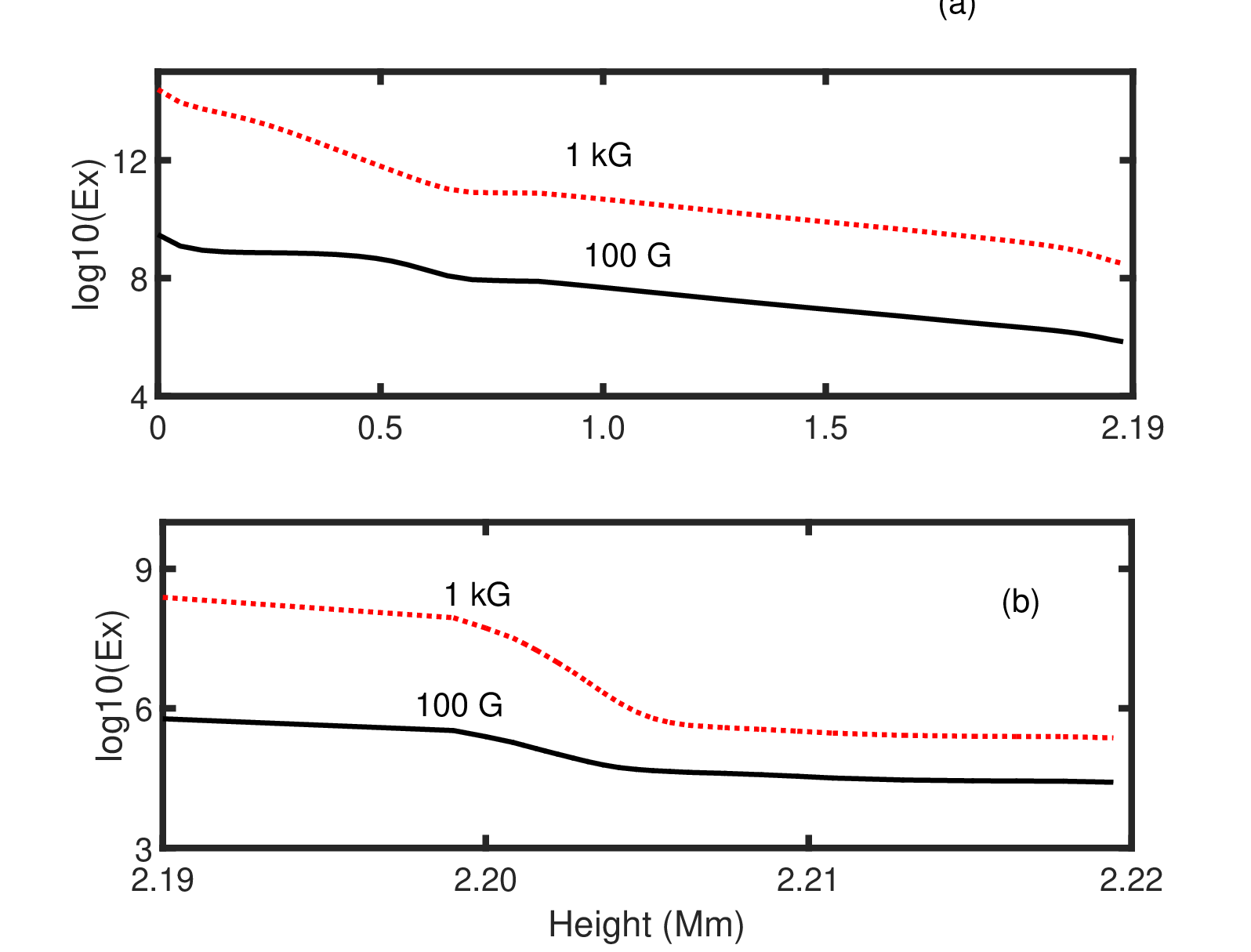}
\caption{The energy flux $log10(Ex)$  is plotted against height for $100 \mbox{G}$ (solid curve) and $5\,\mbox{kG}$ fields (dotted curve).}
\label{fig:F7}  
\end{figure}
To estimate the amount of wave energy that is utilized to heat the plasma, we note that the quantity in the square bracket in Eq.~(\ref{eq:HtR}) is $\sim .01$ for $\delta B/B \sim 0.1$. Thus the heating rate  is $\sim 2\times 10^{-2}\,\Gamma\,\rho\,\va^2$.  Assuming that the bulk fluid velocity $v\sim \va$, the energy flux of the wave 
\bq
Ex=2\times 10^{-2}\,\left(\Gamma\,\rho\,\va^2\right)\,\va\,.
\eq
Solar corona radiative losses are \citep{A07, Y23} $\lesssim 10^5\,\mbox {ergs}\,\persqcm\,\persec$ (quiet regions) and $\sim 10^6-10^8\,\mbox {ergs}\,\persqcm\,\persec$ (active regions). For viable wave heating, chromospheric flux must match these losses ($10^8\,\mbox {ergs}\,\persqcm\,\persec$). Fig.~(\ref{fig:F7}) shows chromospheric/transition region flux greatly exceeds required heating for $\mbox{1\,kG}$ fields. Even with $100\,\mbox{G}$ fields, the flux adequately offsets radiative losses in both regions, supporting wave heating as a viable mechanism for million-degree coronal heating.  

\section{Viscous instabilities, $\lowercase{s}\neq 0$:}

Vortex motions contribute to both chromospheric heating and mass transport. Numerical simulations show viscous heating at vortex sites \citep{M11, K12, Y20, Y21, B21}. Analyzing viscous instabilities provides analytical insight into how shear (vortex) flows affect energy and mass transport between the photosphere and corona. 

 The physical condition of the solar atmosphere allows us to simplify the dispersion relation,  Eq.~(\ref{eq:mE}). For example, in the photosphere and lower chromosphere, $\nu_0\approx \nu_1 \approx \nu_2$ (Fig.~\ref{fig:F1}). In the middle chromosphere, difference between the parallel and perpendicular viscosities are quite small but not zero. Thus, in the following, we shall analyze the  dispersion relation,  Eq.~(\ref{eq:mE}) in the various limiting cases.  

\begin{figure}
\includegraphics[scale=0.34]{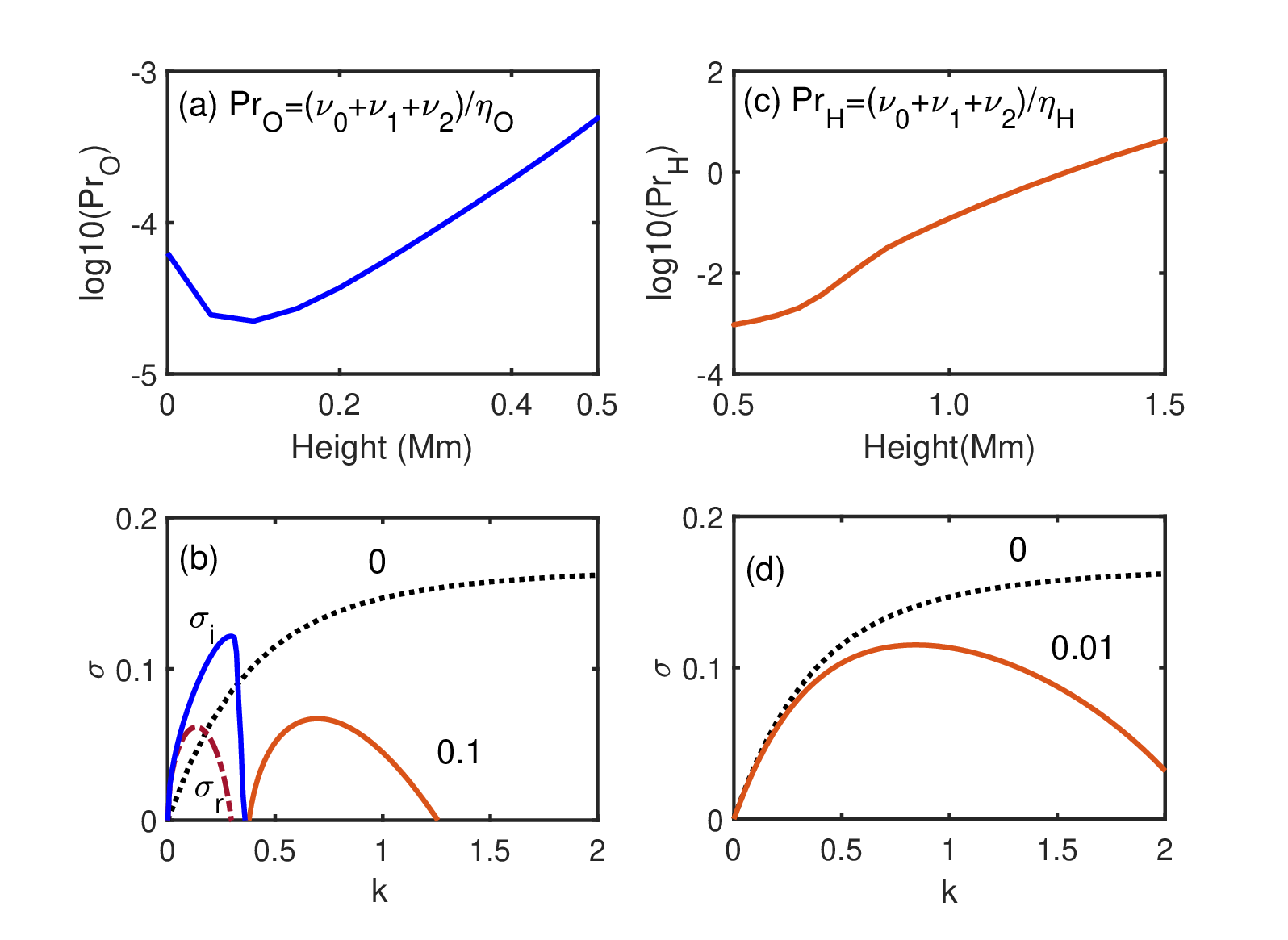}
\caption{The ratio of Ohm ($\eta_O$) and Hall ($\eta_A$) diffusivities to total viscosity ($\nu_0+\nu_1+\nu_2$) is plotted versus height for the photosphere (a) and chromosphere (c). Panel (b) shows viscous instability versus $k$ for $\eta_O\,s/\va^2=0$ and $0.1$, while panel (d) compares $\eta_H\,s/\va^2=0$ and $0.01$. Parameters used: $B_0=100,\mbox{G}$ (c), $\nu_0=1$, $\nu_1=0.98$, $\nu_2=0.99$, $s=10$, $b_z=1$, $k_z=0.4$.}
\label{fig:F9}  
\end{figure}

Viscosity and magnetic diffusivity effects compete in the photosphere-chromosphere region. Using viscosity and magnetic diffusivity values from PW22, we examine their combined impact on wave propagation. The Prandtl number range is wide - in the photosphere, Ohmic diffusion dominates as Ohm Prandtl number, $Pr_{O}=(\nu_0+\nu_1+\nu_2)/\eta_O$ is small [Fig.~\ref{fig:F2}(a), also see below Fig.~~\ref{fig:F9}(a)]. 

In the absence of magnetic diffusion, when the parallel and perpendicular viscosities are equal, i.e., $\nu_0=\nu_1=\nu_2$, wave propagation in the plasma experiences only viscous damping. In this case, the free shear flow energy cannot be transferred to the waves, and thus wave amplification is suppressed. 

However, the presence of Hall and ambipolar diffusion changes this scenario. These non-ideal effects allow the shear flow energy to be transferred to the waves, enabling the growth of fluctuations at all wavelengths, as discussed in [\cite{PW13}; hereafter PW13]. 

Nonetheless, when viscosity is introduced into the system, it counteracts the instabilities driven by Hall and ambipolar diffusion. Viscosity, especially when significant, restricts the Hall and ambipolar instabilities, confining them to long wavelengths. At shorter wavelengths, the viscous damping dominates, preventing the growth of perturbations that would otherwise occur due to the magnetic diffusion effects.

This means that in a regime where Hall and ambipolar diffusion would ordinarily lead to wave growth at a wide range of wavelengths, the presence of viscosity suppresses these instabilities, leaving only the long-wavelength fluctuations to grow. 

Defining viscous frequencies, $\omega_0=k^2\,\nu_0\,,\omega_1=k^2\,\nu_1\,,\omega_2=k^2\,\nu_2\,,\omega_3=k^2\,\nu_3$ and $\omega_4=k^2\,\nu_3$ we consider the following cases.    

{\bf Case I(a):} $\omega_0=\omega_1=\omega_2$, $\eta_H\neq 0\,, \eta_O=\eta_A=\nu_3=0$\\
The dispersion relation Eq. (\ref{eq:mE}) takes the form:
\begin{equation}
\sigma^4 + a_3\sigma^3 + a_2\sigma^2 + a_1\sigma + a_0 = 0,,
\label{eq:DRPH}
\end{equation}
where the coefficients are given by:
\begin{eqnarray}
a_3&=&2\,\omega_2\,,\nonumber\\ 
a_2&=&2\,\left(\mu\,\omA\right)^2+\omega_2^2+\left(s-\omega_{yx}\right)\omega_{xy}\,,\nonumber\\  
a_1&=&2\,\left(s-\omega_{yx}\right)\,\omega_2\,\omega_{xy}\,,\nonumber\\
a_0&=&\left(s-\omega_{yx}\right)\,\omega_2^2\,\omega_{xy}+\left(\mu\,\omA\right)^2\left(s\,\omega_{xy}+\mu^2\,\omA^2\right)\,.
\end{eqnarray}
Here, 
\bq
\omA=k\,\va\,,\mu=\kh\cdot\bb\,,
\label{eq:AFR}
\eq
is the \alf frequency and obliqueness of wave respectively. From the dispersion relation Eq. (\ref{eq:DRPH}), we observe that in the absence of Hall effects [where $\omega_{xy} = H\,k^2\eta_H = 0$ and $\omega_{yx} = -b_z k^2\eta_H = 0$, with $H$ being the helicity, Eq. (\ref{eq:hct})], the shear flow does not couple with viscosity. The energy transfer from free shear flow to waves occurs exclusively through the Hall diffusion of the magnetic field.

Adopting $\va$ and $\nu_0$ as units, we may write $\omA = k$ , $\omega_2 = k^2$  and the necessary condition for the instability becomes
\bq
-s > \frac{b_z\,\left(R_H^2\,k^4+\mu^2\right)}{R_H\,k^2+\mu^2}\,,
\label{eq:nc2}
\eq
where
\bq
R_H=\frac{\eta_H}{\nu_0}\,.
\eq
\begin{figure}
 \includegraphics[scale=0.34]{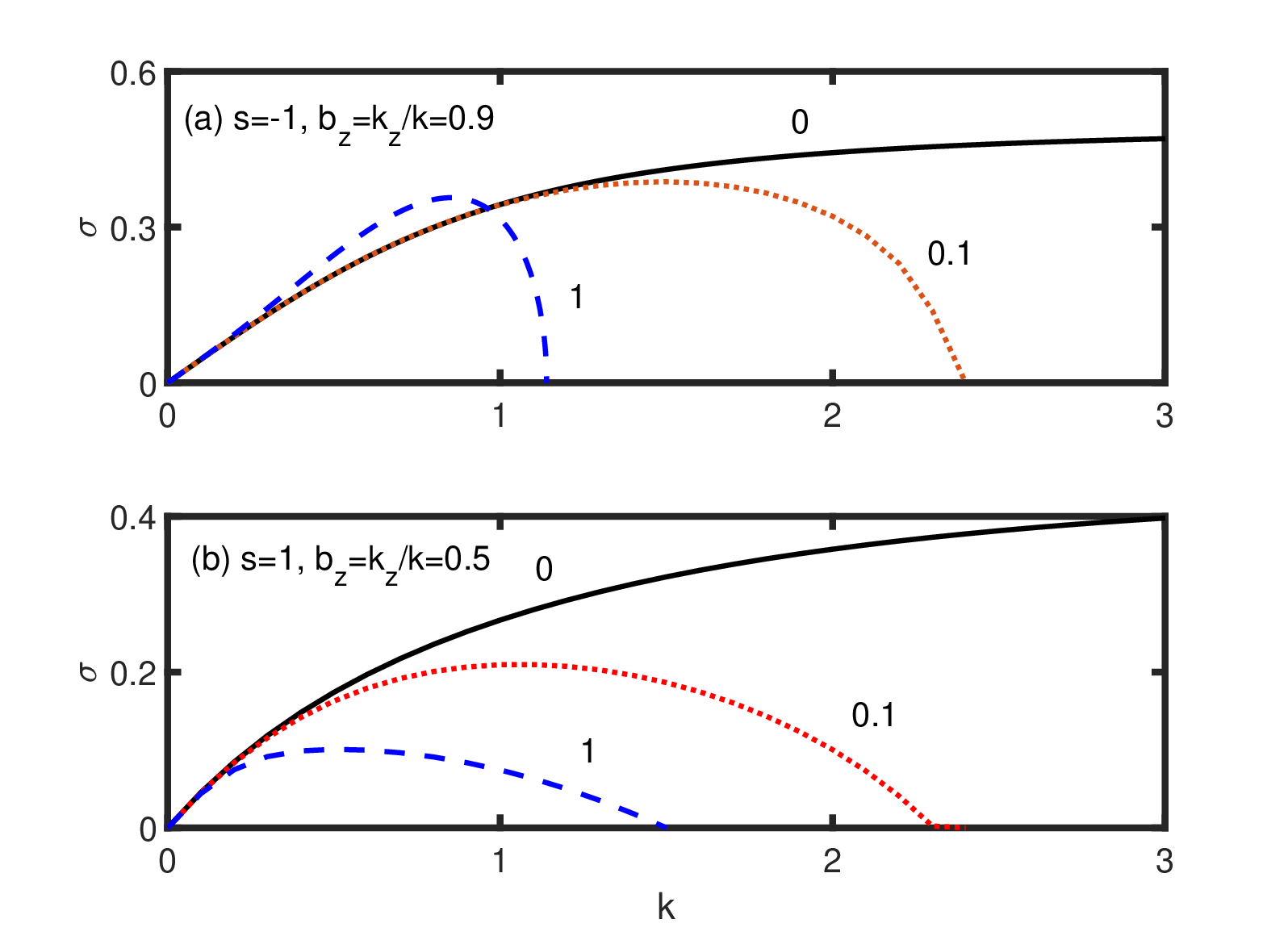}
\caption{The growth rate $\sigma (= \sigma\,\nu_0/\va^2) $ vs. $k (= k\,\va$) is plotted for both the Hall-viscous (panel a) and the ambipolar-viscous (panel b) case. Each curve is labeled according to its respective viscosity value, $\nu_0$. 
The figure also displays values for $s$, $\kz$, and $b_z$. Here, $\kx = \sqrt{1 - \kz^2}$ and $b_y = \sqrt{1 - b_z^2}$, with $s = s , \nu_0 / \va^2$.}
\label{fig:F8}  
\end{figure}
Equation (\ref{eq:nc2}) reduces to Equation (37) of PW13 in the long-wavelength ($k \to 0$) limit. The introduction of viscosity inhibits the Hall instability at small wavelengths, as illustrated in Figure \ref{fig:F8}(a). As the viscosity parameter $\nu_0$ increases from 0 to 1, the instability becomes increasingly confined to long-wavelength fluctuations.
In the absence of viscosity, short-wavelength (large $k$) fluctuations which were growing at a constant rate due to Hall diffusion, as given by Equation (52) of PW13:
\begin{equation}
\sigma = \sqrt{\kz^2\, b_z(-s-b_z)},.
\label{eq:growth_rate}
\end{equation}
are now subject to viscous damping. This demonstrates how viscosity acts as a stabilizing mechanism at small scales, while preserving the long-wavelength Hall instability.

{\bf Case I(b): $\omega_0=\omega_1=\omega_2$, $\eta_A\neq 0$\,, \bf $\eta_O=\eta_H=\nu_3=0$}\\
In this case, aside from the coefficients, the dispersion relation is identical to Eq.~(\ref{eq:DRPH}). Defining
\bq
\omega_{AD}=k^2\,\eta_A\,,\quad U=g\,s+\mu^2\,\omD\,,
\eq
the coefficients are
\begin{eqnarray}
a_3&=&2\,\omega_2+\left(1+\mu^2\right)\,\omD\,,
\nonumber\\
a_2&=&2\,\mu^2\,\omA^2 + \omega_2^2 + \Big[U+2\left(1+\mu^2\right)\omega_2\Big]\,\omD 
 \,,  
\nonumber\\
a_1&=&2\,\mu^2\omA^2\omega_2+\Big[2\,\omega_2\,U+\left(1+\mu^2\right)\left(\omega_2^2+\mu^2\,\omA^2\right)\Big]\omD
\nonumber\\
a_0&=&
U\,\omega_2^2\,\omD
+
\Big[\mu^2\,\omA^2+\left(g\,s+\left(1+\mu^2\right)\omega_2\right)\,\omD
\Big]\mu^2\,\omA^2
\nonumber\\
\,.
\label{eq:DRP3}
\end{eqnarray}
Here,
\bq 
g=-\kx\,\kz\,b_y\,b_z\,,
\eq
is the topological switch.
As in the previous case, in the absence of ambipolar diffusion, shear does not couple with viscosity, and the coefficients above match those in Eq.~(24) of PW13. The necessary condition of the instability, $a_0<0$  becomes
\bq
 s>\frac{\kz\,b_z}{\kx\,b_y}\,\frac{\mu^2+\left(1+\mu^2+k^2\,R_A\right)k^2\,R_A}{R_A\left(\mu^2+k^2\right)}\,,  
\label{eq:ncA}
\eq
Here
\bq
R_A=\frac{\eta_A}{\nu_2}\,.
\eq
In the absence of viscosity, (i.e. setting $k=0$), Eq.~(\ref{eq:ncA}) reduces to Eq.~(34) of PW13 which is the necessary condition for the ambipolar instability. Growth rate of the instability in the $k\rightarrow \infty$ limit is
\bq
\sigma=\frac{1}{2}\Big[-\left(1+\mu^2\right)\pm\sqrt{\left(1+\mu^2\right)^2-4\,g\,s\,R_A}\Big]\,.
\eq
The expression presented corresponds to Equation (54) of PW13. However, it is important to note a typographical error in their original equation: within the square root term, $(1-\mu^2)^2$ should be corrected to $(1+\mu^2)^2$.

Analogous to the Hall case, viscosity acts to suppress the onset of ambipolar instability, with this effect being particularly pronounced at short wavelengths. As illustrated in Fig.~[\ref{fig:F8}(b)], increasing the viscosity parameter from $0$ to $1$ (with curves labeled according to their respective viscosity values) results in growth being restricted to long-wavelength fluctuations only. However, a notable distinction from the Hall case shown in Fig.~[\ref{fig:F8}(a)] shear gradient in the ambipolar case is positive. 

{\bf Case II:} $\omega_0\neq\omega_1\neq\omega_2$, $\omega_3=0$\\
We shall assume that $\nu_0$ and $\nu_1\,,\nu_2$ are not identical but differ by a very small amount. In this case, perpendicular viscosities may destabilise the wave if the necessary condition, $C_0<0$, or 
\bq
\Big[(\mu\,\omA)^2-G_2(s)\Big](\mu\,\omA)^2<-G_0(s)\,,
\label{eq:ncX}
\eq
is satisfied. Here coefficients $C_0\,,G_2(s)$ and $G_0(s)$ are defined in Eqs.~(\ref{eq:CoE1})) and  Eqs.~(\ref{eq:GoE1})) respectively.

When the magnetic field is purely vertical and wavevector is parallel to the magnetic field, i.e. when $\mu=1$ we have 
$X_1=Y_2=\omega_2\,,S_{xy}=S_{yx}=\Delta_1\,,X_2=Y_1=S_{xx}=S_{yy}=G_2(s)=0\,,
Z_1=-\omega_2\,\Delta_1\,,Z_2=-\Delta_1^2\,,$ and $G_0(s)=-\omega_1\,\Delta_1\,s^2$. In this case, Eq.~(\ref{eq:ncX}) simplifies to 
\bq
s^2>\frac{\omA^4}{\omega_1\,\Delta_1}\,,    
\eq
which corresponds to Eq.~(5) in PW23, excluding the effects of ambipolar diffusion. Given that $\nu_1<\nu_2$ in the solar atmosphere, we have $\Delta_1\equiv \omega_1-\omega_2<0$. 
This leads to a negative right-hand side in the inequality, indicating that even a small difference between the perpendicular viscosities destabilizes the \alf wave. For cases where the magnetic field is not purely vertical and the wave is not purely \alfc, this slight disparity between $\nu_1$ and $\nu_2$ also induces wave instability due to differential damping along $\nu_1$ and $\nu_2$.

In Fig.~\ref{fig:F9}(b), for a vertical field ($b_z=1, k_z=0.4$) with viscosities $\nu_0=1\,,\nu_1=0.98\,,\nu_2=0.99$ and shear $s=10$, the viscous instability growth rate (curve $0$) increases with $k$, peaks, then plateaus.
Adding Ohmic diffusion ($\eta_O=0.1$) limits growth to small $k$ values. At $k\rightarrow 0$, modes become overstable ($\sigma=\sigma_r+i\,\sigma_i$) with small growth rate $\sigma_r$ and large oscillatory component $\sigma_i$. Higher diffusion fully damps the instability. Given the low Prandtl number shown in Fig.~\ref{fig:F9}(a) the viscous instability likely cannot survive in the photosphere.

In Fig.~\ref{fig:F9}(d), the viscous instability growth rate without Hall/Ohmic diffusion (curve $0$) matches Fig.~\ref{fig:F9}(b). With Hall diffusion ($\etaH=0.01$), growth remains purely growing but confined to long wavelengths, similar to the Ohmic case. For this configuration (vertical field, $\kx\,,\kz$ wavevectors, positive shear $s$), Hall diffusion causes dissipationless diffusion rather than destabilization (PW13 Eq. 34). While small viscous momentum diffusion differences drive the instability, Hall diffusion reduces these differences. Like the Ohmic case, large $\etaH$ eliminates the instability. Per Fig.~\ref{fig:F9}(c), the viscous instability may operate in the middle chromosphere ($\sim 1.5\,\mbox{Mm}$), depending on shear gradient sign, when Hall instability is absent.

Given $\etaH\,s/\va^2=0.01$ and $s\,\nu_0/\va^2=10$, we find with $\nu_0+\nu_1+\nu_2\approx 3$, $\nu_0/\etaH\approx 10^2$. Comparing with Fig.~\ref{fig:F9}(c), the viscous instability remains unchanged in the chromosphere [curve $0$ in (d)], only showing effects of Hall diffusion in the transition region [curve $0.01$ in Fig.~\ref{fig:F9}(d)].

 \begin{figure}
\includegraphics[scale=0.34]{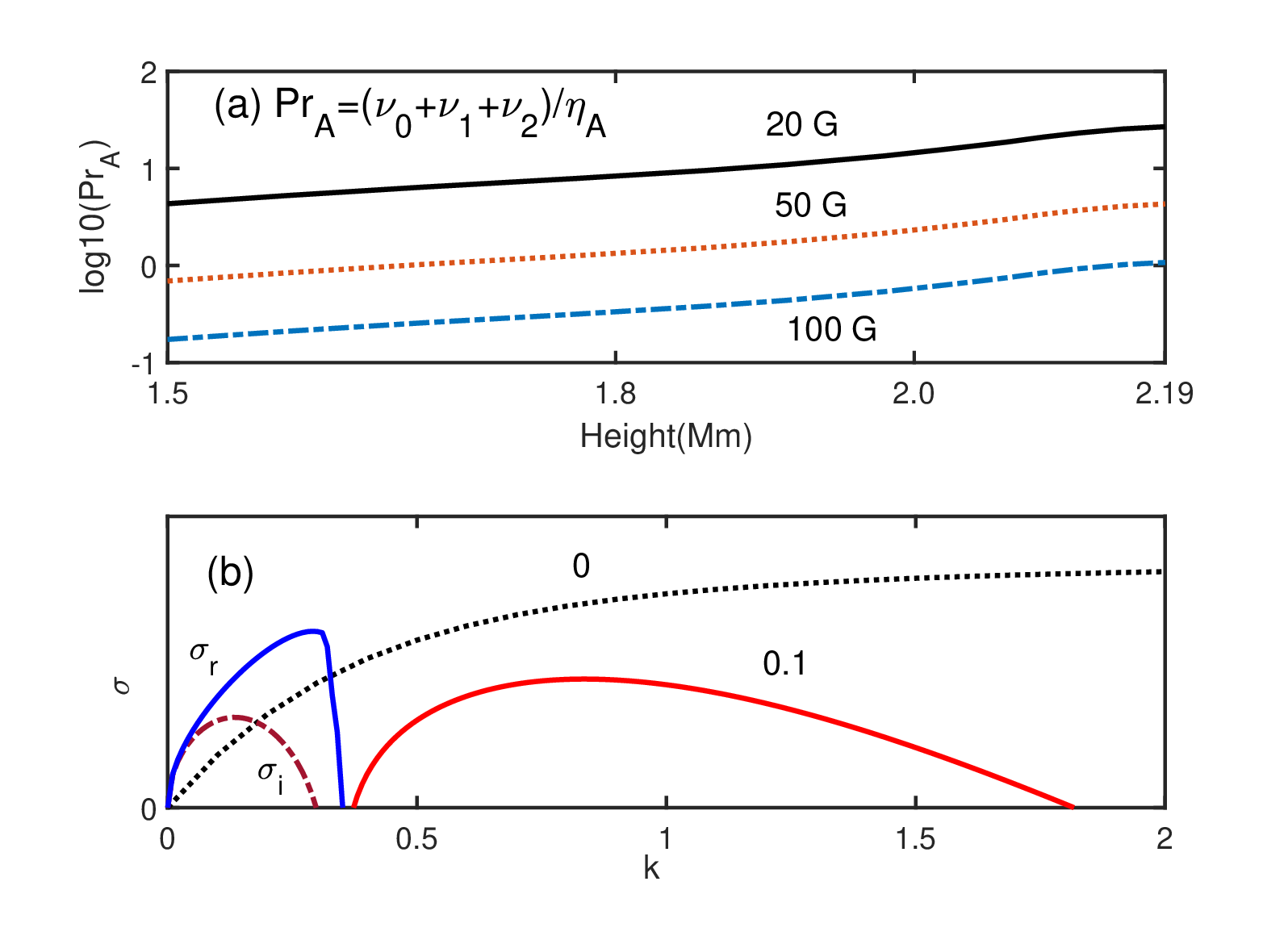}
\caption{The ratio of ambipolar diffusion ($\eta_A$) to total viscosity ($\nu_0+\nu_1+\nu_2$) is plotted for the middle and upper chromosphere (top panels). The lower panel shows growth rate versus $k$ for varying viscosities and ambipolar diffusion. Other parameters remain unchanged from the previous figure.}
\label{fig:F10}  
\end{figure}

The relative importance of ambipolar diffusion, $\eta_A$, compared to combined viscosity $\nu_0+\nu_1+\nu_2$, shown in the top panel of Fig.~\ref{fig:F10}(a) for the middle and upper chromosphere, varies with magnetic field strength. For weak fields ($\lesssim 20\,$ G), viscosity dominates across the chromosphere, while for moderate fields ($\sim 50\,$ G), viscous transport and ambipolar diffusion are comparable. For stronger fields ($\gtrsim 100\,$ G), ambipolar diffusion becomes dominant. These different regimes warrant exploring viscous instability growth under varying Prandtl numbers to assess stability across field strengths.

 The lower panel compares growth rates for viscous instability alone ($\eta_A=0$, curve $0$) and viscous plus ambipolar diffusion ($\eta_A=0.1$, curve $0.1$). Ambipolar diffusion restricts growth to long wavelengths with reduced rates. Very long wavelengths ($k\rightarrow 0$) are overstable, similar to Ohm's case. At $\eta_A=1$ (not shown in the figure), only overstable modes persist. Given varying ambipolar diffusion in the chromosphere, both overstable and unstable modes likely coexist across different wavelengths.

For $\etaA\,s/\va^2=0.1$ and $s\,\nu_0/\va^2=10$, we find $\nu_0/\etaA\approx 10$. From Fig.~\ref{fig:F10}(a), weak fields ($B_0 \sim 20 \mbox{G}$) show viscous instability in the chromosphere [curve $0$ in (b)]. Stronger fields ($B_0 \gtrsim 50 \mbox{G}$) lead to unstable and overstable waves at various wavelengths [curve $0.1$ in (b)] in the upper chromosphere and transition region.

{\bf{Case III:} only $\nu_0$ and  $\eta_H$ are present.}\\
Defining 
\bq
F=1-\mu^2\,,
\label{eq:deF}
\eq
the dispersion relation Eq.~(\ref{eq:mE})  becomes
\begin{eqnarray}
\sigma^4+3\,F\,\mu^2\,\omega_0\sigma^3 + \Big[2\,\left(\mu\,\omA\right)^2+\left(s-\omega_{yx}\right)\omega_{xy}\Big]\sigma^2  
\nonumber\\
+\Big[3\,F\,\mu^2\,\omega_0\,\omA^2 + 3\left(s\,\kx^2\,\bz2-F\,\omega_{yx}\right)\omega_0\,\omega_{xy}\Big]\mu^2\,\sigma
\nonumber\\
+\Big[\left(\mu\,\omA\right)^2+s\,\omega_{xy}\left(1-3\,g\,b_z\,\frac{\eta_H\,\omega_0}{\va^2}\right)\Big]\left(\mu\,\omA\right)^2=0\,.
\label{eq:DRP}
\end{eqnarray}
Here
$\omega_{yx}=-b_z\,k^2\,\eta_H$ and $\omega_{xy}=H\,k^2\,\eta_H\,.$

We observe from the dispersion relation that when Hall effects are absent ($\eta_H = 0$ or equivalently $\omega_{xy} = \omega_{yx} = 0$), there is no coupling between shear flow and parallel viscosity. This suggests that parallel viscosity can only act to damp wave propagation.
Setting $\eta_H = 0$, the dispersion relation reduces to:
\bq
\left(\sigma^2+\mu^2\omA^2\right)
\Big[\sigma^2+3\,\left(1-\mu^2\right)\mu^2\,\omega_0\,\sigma+\mu^2\omA^2\Big]=0\,,
\label{eq:disp_no_hall}
\eq
This equation yields distinct behaviors in two cases: (i)  When $b_z = 1$ and $k_z = 1$ (implying $\mu = 1$), Eq. (\ref{eq:disp_no_hall}) describes undamped \alf waves. In this configuration, parallel viscosity has no effect on transverse fluctuations; (ii) When the magnetic field has both vertical and azimuthal components ($\mu \neq 1$), we obtain two sets of solutions: (a) Undamped \alf waves with frequency $\sigma = \pm i\mu\omega_A$ (b) Viscously damped waves with the damping rate for the root
\bq
\frac{\sigma}{\mu\,\omA}=-\frac{3}{2}\left(1-\mu^2\right)\mu\,\frac{\omega_0}{\omA}\pm i\Big[1-\frac{3}{2}\left(1-\mu^2\right)^2\mu^2\,\frac{\omega_0^2}{\omA^2}\Big]^{1/2}\,,
\eq
is
\bq
\gamma=\frac{3}{2}\left(1-\mu^2\right)\mu^2\,\omega_0\,.
\eq
To summarize, parallel viscosity damps the wave propagation in the medium only if the magnetic field has both the vertical and azimuthal components and  the wavevector has parallel and transverse components. This demonstrates the geometric nature of viscous damping in magnetized plasmas.

When both Hall effects and parallel viscosity are present, wave instability occurs when:
\begin{equation}
b_z v_A^2 < -s\eta_H\left(1-3gb_z\frac{\eta_H\omega_0}{v_A^2}\right)
\label{eq:NCP}
\end{equation}
This necessary condition reveals that the combination of Hall diffusion of the magnetic field and parallel viscous momentum transport can redirect free shear energy into wave growth.

Two limiting cases emerge: (1) In the absence of parallel viscosity ($\omega_0=0$), Eq. (\ref{eq:NCP}) reduces to:
\begin{equation}
-s > b_z\omega_H
\label{eq:HI}
\end{equation}
This recovers Equation (37) of PW13—the necessary condition for Hall instability. Here we have used  $\eta_H=\va^2/\omega_H$ where $\omega_H=(\rho_i/\rho_n)\,\omega_{ci}$ \citep{PW08}.

(2) Strong Viscous-Hall Coupling ($3gb_z\eta_H\omega_0/v_A^2 \gg 1$)
The necessary condition becomes:
\begin{equation}
s > \frac{1}{3g}\frac{\omega_H^2}{\omega_0}
\label{eq:VHI}
\end{equation}
This defines a new regime we term the "viscous-Hall instability."

The sign of the shear flow gradient determines which instability mechanism dominates: (i) Negative shear gradients tend to trigger the Hall instability. (ii) Positive shear gradients can drive the viscous-Hall instability, even when conditions are unfavorable for pure Hall instability
This demonstrates how the interplay between Hall effects and viscosity creates new pathways for instability that are absent in either pure Hall or pure viscous regimes.

Adopting $\va$ and $\nu_0$ as units, we may write $\omA = k$ , $\omega_0 = k^2$ and the dispersion relation, Eq.~(\ref{eq:DRP}) becomes
\begin{eqnarray}
\sigma^4+\Big[\left(3\,F\,\sigma+R_H^2\,k^2+2\right)\,\mu^2 + s\,R_H\,\kz^2\,b_z\Big]\,k^2\sigma^2 + 
\nonumber\\
3\,\mu^4\,k^4\,\Big[\left(1+R_H^2\,k^2\right)\,F + s\,R_H\,\kx^2\,b_z\Big]\,\sigma
+
\nonumber\\
\mu^2\,k^4\,\Big[\left(1-3\,s\,g\,R_H^2\,k^2\right)\mu^2+s\,R_H\,\kz^2\,b_z\Big]=0\,.
\label{eq:DRP1}
\end{eqnarray}
In the limiting case $k\rightarrow 0$ ($\sigma\sim k$), the growth rate of the viscous-Hall instability becomes
\bq
\sigma=k\,\mu\,\sqrt{\Big|1+R_H\,\frac{s}{b_z}\Big|}\,,
\eq
when $1+R_H\,s/b_z<0$.
\begin{figure}
 \includegraphics[scale=0.34]{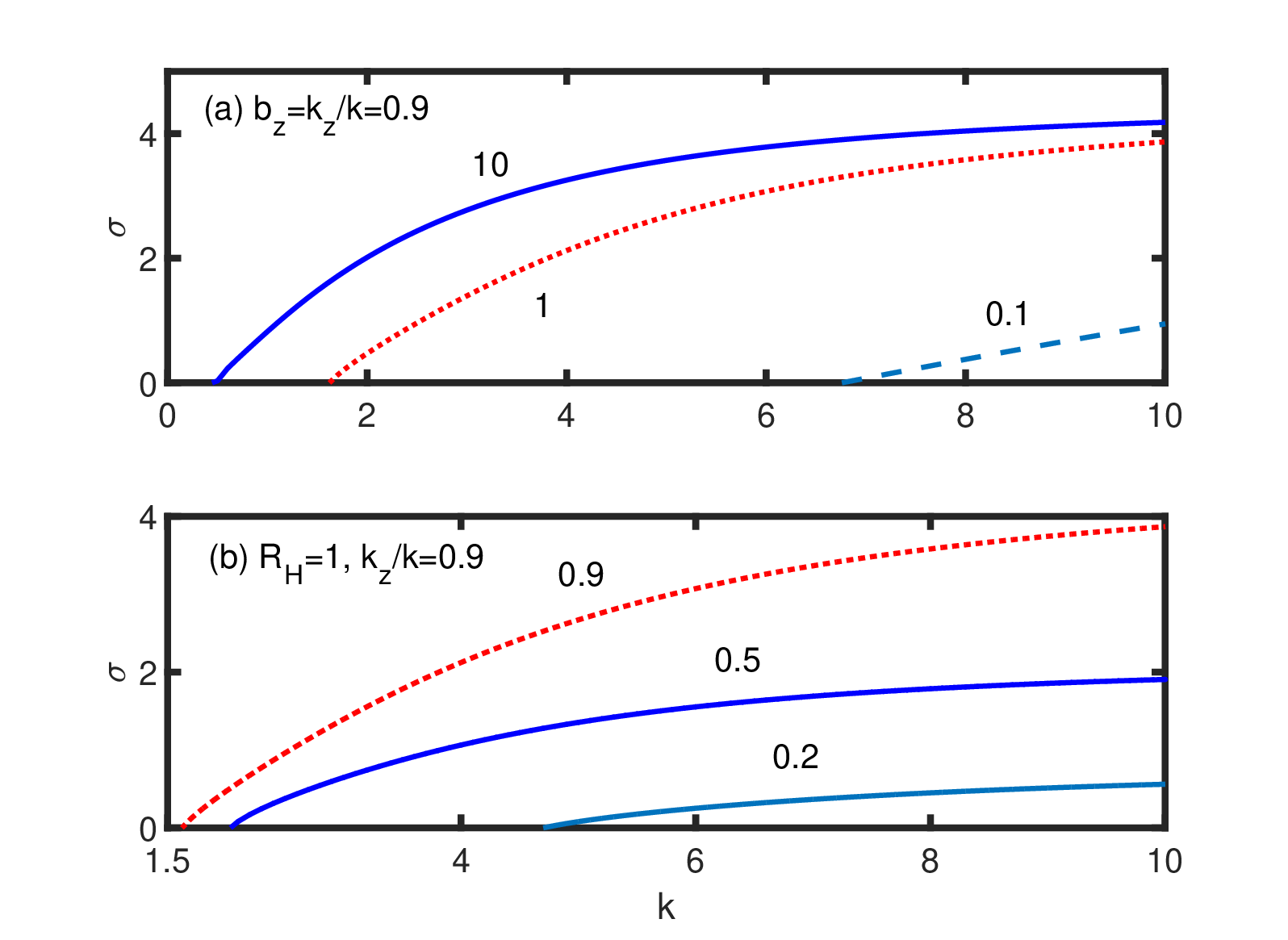}
\caption{The growth rate $\sigma$ vs. $k$ is plotted for $s = 10$ and $b_y = -\sqrt{1 - b_z^2}$. In panel (a), $R_H = \eta_H / \nu_0$ (labeled on the curve) is varied for $\kz = b_z = 0.9$. In panel (b), with $R_H = 1$ and $\kz = 0.9$, the value of $b_z$ (also labeled on the curve) is varied.} 
\label{fig:F11}  
\end{figure}

The dispersion relation, Eq.~(\ref{eq:DRP1}) is solved numerically for $s=10$ and $b_y=-\sqrt{1-\bz2}$. The results are shown in Fig.~(\ref{fig:F11}) where in the top panel (a) the  values of $b_z$ and $\kz$ is kept fixed and the ratio of Hall diffusion and parallel viscosity, $R_H$ is varied while in the bottom panel (b) the values of $R_H$ and $\kz$ is kept fixed and the value of $b_z$ is varied. We see from panel (a) that with the decreasing $R_H$, i.e with the increasing strength of the parallel viscosity, it is only the short wavelength fluctuations that are subject to the viscous-Hall instability with considerably reduced growth rate. The long wavelength fluctuations are damped in this case. Clearly, larger values of Hall diffusion yields bigger growth rate for the 
viscous-Hall instability.  Notably, neither Hall nor parallel viscosity alone can channel free shear energy to waves—both magnetic diffusion and viscous momentum transport must operate together.

As can be seen from  Fig.~\ref{fig:F11}(b), the most favourable magnetic topology for the viscous-Hall instability is when the field is almost vertical ($b_z=0.9$) and waves are propagating almost parallel ($\kz=0.9$) to the field. With the increase in the azimuthal ($-b_y$) field, the instability grows at a much reduced rate and that too only for the small wavelength fluctuations. As is clear from Fig.~\ref{fig:F9}(c), in the lower and middle chromosphere, $R_H\sim 1/Pr_H<1$ and in the upper chromosphere  $R_H\sim 1$. Thus, the viscous-Hall instability may operate across the whole chromosphere with the growth rate depending on the exact value of $R_H$.

{\bf{Case IV:} only $\nu_0$ and  $\eta_A$ are present.}\\
The dispersion relation in Eq. (25) in this case has the following coefficients in units of $\va$ and $\nu_0$
\begin{eqnarray}
\frac{C_3+E_3}{k^2}&=&3\,\mu^2\,F+\left(1+\mu^2\right)\,R_A\,, 
\nonumber\\
\frac{C_2+E_2}{\mu^2\,k^2}&=&2+k^2\,R_A\,\big[3\,F\,\left(1+\mu^2\right)+R_A]+\frac{s\,g}{\mu^2}\,R_A\,,
\nonumber\\
\frac{C_1+E_1}{\mu^2\,k^4}&=&\frac{C_3+E_3}{k^2}+3\,\mu^2\left(F\,k^2\,R_A-s\,g\right)\,R_A\,,
\nonumber\\
\frac{C_0+E_0}{\mu^4\,k^4}&=&1+\frac{s\,g}{\mu^2}\,R_A+3\,k^2\,R_A\,\left(F-s\,g\,R_A\right)\,.
\label{eq:DRA}
\end{eqnarray}
Here $F$ is defined in Eq.~(\ref{eq:deF}) while the coefficients $C_j$ and $E_j$ with $j=0\,,1\,,2\,,3$ in Eqs.~(\ref{eq:CoE1}) and  Eqs.~(\ref{eq:deE}) in the appendix.

While parallel viscosity alone only damps waves, its combination with ambipolar diffusion, similar to Hall effects, can trigger instability. The necessary condition for the viscous-ambipolar instability, $C_0+E_0<0$ becomes
\bq
-s\,g\,R_A>\mu^2\,\left(1+\frac{3\,k^2\,R_A}{1-3\,\mu^2\,R_A\,k^2}\right)\,,
\eq
from which the necessary condition for the ambipolar instability, $-s\,g\,\eta_A>\mu^2$ [Eq.~(36)] of PW13 is recovered in the long wavelength ($k\ll 1/\mu\sqrt{3\,R_A}$) limit.  The growth rate of the viscous-ambipolar instability in this limit becomes
\bq
\sigma=k\,\mu\,\sqrt{-s\,g\,R_A-\mu^2}\,,
\eq
which for $-s\,g\,R_A>\mu^2$ is real.

On the other hand, in the $k\gg 1/\mu\sqrt{3\,R_A}$ limit, the above condition becomes
\bq
s\lesssim \frac{F}{g\,R_A}\,.
\eq
Since the geometric factor $F/g$ is order unity, the viscous-ambipolar instability depends  solely on the ratio $1/R_A=\nu_0/\eta_A$. Numerical solutions of the dispersion relation reveal that this results in an overstable mode, with growth rates significantly lower than those of the pure ambipolar instability.

\begin{figure}
 \includegraphics[scale=0.34]{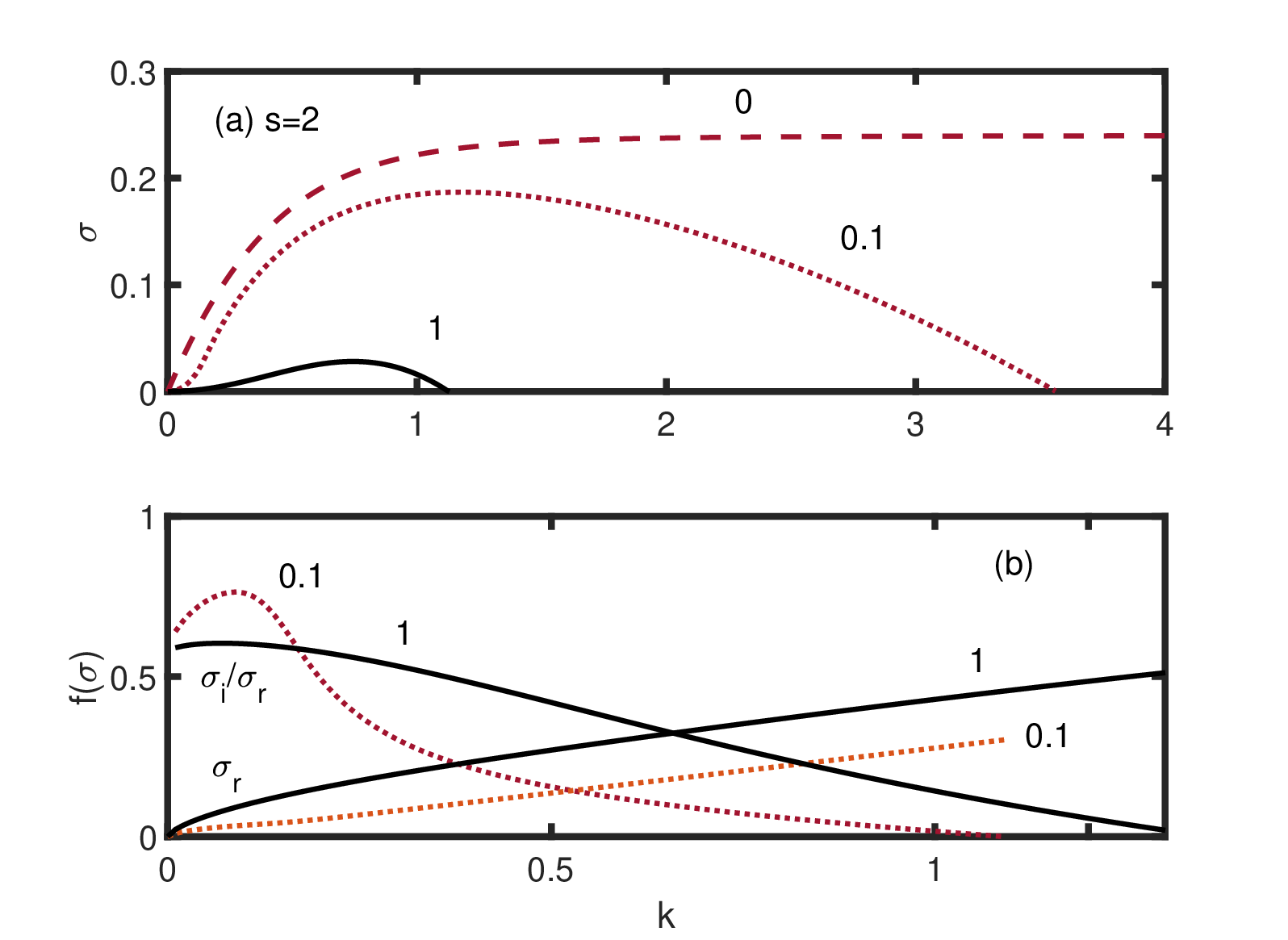}
\caption{The growth rate $\sigma$ vs. $k$ is plotted for  $s=2$ and $b_z=\kz=1/2$. In the panel (a) purely growing mode is  plotted for varying $R_A$ and in the 
(b) corresponding overstable mode is plotted.} 
\label{fig:F12}  
\end{figure}

Fig.~\ref{fig:F12} shows the dispersion relation (Eq.~\ref{eq:mE}) with coefficients from Eq.~(\ref{eq:DRA}), for $s=2$ and $b_z=\kz=0.5$. Panel (a) demonstrates that while the purely growing ambipolar instability exists when $\nu_0=0$, introducing viscosity ($\nu_0=0.1\,,1$) constrains the instability to longer wavelengths and reduces growth rates. At high viscosity, this mode may vanish entirely. Panel (b) reveals that viscosity induces an overstable mode, where $\sigma_i\lesssim \sigma_r$ for $k\ll 0.5$, indicating overstability in long-wavelength fluctuations. The imaginary frequency $\sigma_i$ decreases rapidly with increasing $k$. Thus, viscosity both restricts the ambipolar instability to longer wavelengths with reduced growth rates and generates a weak overstable mode at shorter wavelengths.

{\bf Case V. Only gyroviscosity is present, i.e.  $\omega_3\neq0$ and $\omega_4\neq0$}.\\
The dispersion relation, Eq.~(\ref{eq:mE}) becomes
\bq
\sigma^4+C_2\,\sigma^2+ C_1\,\sigma + C_0=0\,,
\label{eq:dr3}
\eq
where
\begin{eqnarray}
C_2=\mu^2\Big[2\,\,\omA^2+\left(F\,\Delta_3+\mu^2\,\omega_4\right)^2\Big]
\nonumber\\
-s\,\mu\,\kz\,\Big[3\,\by2\,\left(\Delta_3-\omega_4\right)+\omega_4\Big]\,,
\nonumber\\
C_1=-3\,\mu^2\,g\,s\,\left(F\,\Delta_3+\mu^2\,\omega_4\right)\left(\Delta_3-\omega_4\right)\,,
\nonumber\\
\frac{C_0}{\mu^2}=\omA^2\,\Big[\left(\mu\,\omA\right)^2-\mu\,\kz\,s\left(3\,\by2\,\left(\Delta_3-\omega_4\right)+\omega_4\right)\Big]
\nonumber\\
-s^2\Bigl\{\kz^2\Big[\left(1+F\right)\Delta_3+\mu^2\,\omega_4-\kx^2\bz2\left(\Delta_3-\omega_4\right)\Big]\times
\nonumber\\
\Big[\left(\bz2-\by2\right)\,\Delta_3+2\,\by2\,\omega_4\Big]-2\,g^2\,\left(\Delta_3-\omega_4\right)^2
\Bigr\}\,.
\label{eq:coef1}
\end{eqnarray}
Here for $\omega_3=\alpha\,\omega_4$ where $\alpha=\nu_3/\nu_4$ (Eq.~\ref{eq:alp}),
\bq
\Delta_3=\omega_3-\omega_4\equiv \left(\alpha-1\right)\omega_4\,.    
\eq
The above dispersion relation, Eq.~(\ref{eq:dr3}) reduces to Eq.~(57) of PW22 for $\alpha=1/2$ and $\mu=1$. 

In the units of $\va$ and $\nu_4$ above coefficients, Eq.~(\ref{eq:coef1}) becomes 
\begin{eqnarray}
\frac{C_2}{\mu^2\,k^2}&=&2
- \frac{s}{b_z}\,q_1 + k^2\,\left(\alpha_1-\mu^2\,\alpha_2\right)^2\,,
\nonumber\\
\frac{C_1}{\mu^3\,k^3}&=&3\,b_y\,\kx\,k\,s\,\alpha_2\,\left(\alpha_1-\mu^2\,\alpha_2\right)\,,
\nonumber\\
\frac{C_0}{\mu^4k^4}&=&1-\frac{s}{b_z}\,q_1
-\frac{s^2}{\bz2}\,q_2
\,.
\nonumber\\
\label{eq:coef2}
\end{eqnarray}
Here 
\bq
\alpha_1=\alpha-1\,,\quad \alpha_2=\alpha-2\,.
\eq
and
\begin{eqnarray}
q_1&=&\left(1+3\,\alpha_2\,\by2\right)\,,\nonumber\\
q_2&=&\alpha\,\alpha_1\,\bz2+2\,\by2\left(1-F\,\alpha_2^2\right)\,.
\end{eqnarray}
\begin{figure}
\includegraphics[scale=0.34]{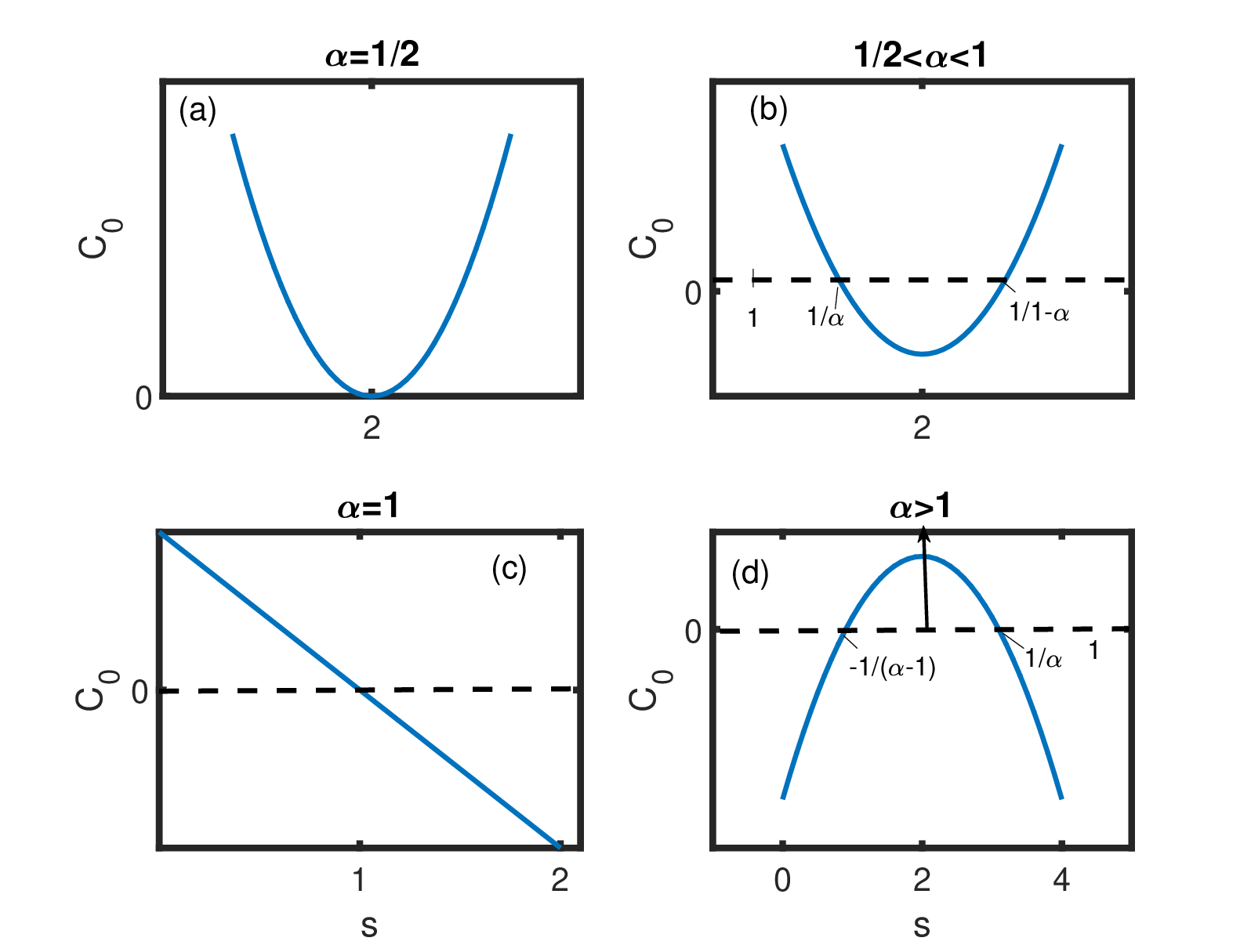}
\caption{The coefficient $C_0$ is plotted against $s$ for fixed $\alpha$ in the above figure\,.}
\label{fig:F13}  
\end{figure}
For $\sigma_1=\sigma/\mu\,k$ the dispersion relation, Eq.~(\ref{eq:dr3}) becomes
\bq
\sigma_1^4+\frac{C_2}{\mu^2\,k^2}\,\sigma_1^2+\frac{C_1}{\mu^3\,k^3}\,\sigma_1+\frac{C_0}{\mu^4k^4}=0\,.
\label{eq:dr4}
\eq
Given the complexity of the dispersion relation, Eq. (\ref{eq:dr4}), we first analyze a simplified case to gain analytical insight. We consider: $\mu = 1$ (purely vertical field) and $k_z = 1$ (parallel propagation). While PW22 analyzed this configuration specifically for $\alpha = 2$, we extend the analysis to various values of $\alpha$, reflecting the height-dependent variation of this parameter in the solar atmosphere [Fig.~\ref{fig:F4}].\\ 

{\bf{Special case: $b_z=1\,,\kz=1$}}\\

In this case, the dispersion relation  Eq.~(\ref{eq:dr3}) becomes
\bq
\frac{\sigma^4}{k^4}+C_2\,\frac{\sigma^2}{k^2}+C_0=0\,.
\label{eq:disx}
\eq
Here
\begin{eqnarray}
C_2 &=& 2 - s + k^2\,,\nonumber\\
C_0 &=& \left(\alpha\,s-1\right)
\left(\left(1-\alpha\right)\,s-1\right)\,.
\label{eq:cof1}
\end{eqnarray}

The discriminant, $D=C_2^2-4\,C_0$ of  Eq.~(\ref{eq:disx}) is
\bq
D=\left(k^2+2-s\right)^2-4\,C_0\equiv \left(k^2-k_1^2\right)\left(k^2-k_2^2\right)\,,
\label{eq:dis1}
\eq
with
\bq
k_{1\,,2}^2=s-2\pm2\,\sqrt{C_0}\,.
\eq
As $4\,C_0=\left(s-2\right)^2-\left(2\,\alpha-1\right)^2\,s^2$, we get $2\sqrt{C_0}<|s-2|$. Therefore, $k_1^2$ and $k_2^2$ have same sign.

Note that when $D>0$, we have two real, $\sigma_{\pm}^2$ roots, while $D<0$ implies pair of complex 
conjugate roots, $\sigma^2=\left(\gamma\pm i\,\omega\right)^2$. For fixed $\alpha$ the value of shear $s$ determines whether $C_0$ is positive, or negative, i.e. whether waves are stable, overstable, or unstable. We see from Fig.~(\ref{fig:F13}) that the value of $C_0$ changes with $\alpha$ and $s$. In general following possibilities exist:\\
\begin{figure}
\includegraphics[scale=0.34]{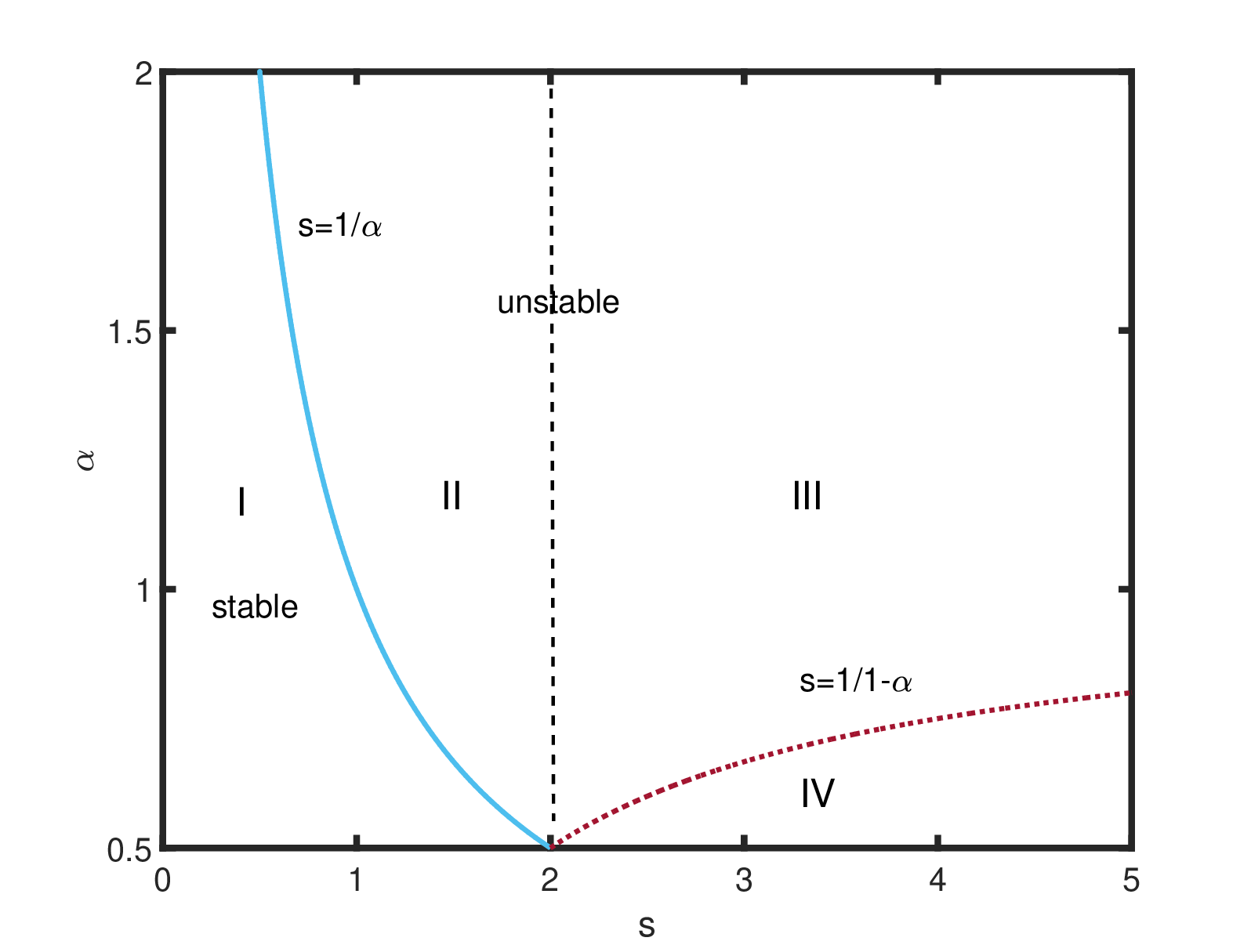}
\caption{In the above figure, the unstable and stable regions is delineated using the necessary condition, Eq.~(\ref{eq:nc1}).}
\label{fig:F14}  
\end{figure}
{\bf {1. $C_0<0\, (D>0)$:}} One positive and one negative  root, i.e. 
\bq\sigma^2= -\omega^2\,,\gamma^2\,,\eq
and we have purely growing mode with the growth rate $\sim \gamma\,.$\\

From Eq.~(\ref{eq:cof1}) we see that $C_0<0$ if
\bq
s \in\left(\frac{1}{\alpha}\,,\,\,\frac{1}{1-\alpha}\right)\, \text{if} \,\,\frac{1}{2}<\alpha<1\,.                
\label{eq:nc1}
\eq
This is seen in Fig.~\ref{fig:F13}(b). Thus, the waves are unstable if they belong to the region II and III in Fig.~(\ref{fig:F14}). Writing the dispersion relation as $a_2\,k^4+b_2\,k^2+c_2=0$ where
\bq
a_2=C_0+\sigma^2\,,\quad b_2=\left(2-s\right)\sigma^2\,,\quad C_2=\sigma^4\,,
\eq
we see that the discriminant $D_2=\Big[\left(2\,\alpha-1\right)^2\,s^2-4\,\sigma^2\Big]\sigma^4$ is positive only if 
$\sigma^2<\left(\alpha-1/2\right)^2\,s^2$. Thus the maximum attainable growth rate of the instability
\bq
\sigma_0^2=\left(\alpha-1/2\right)^2\,s^2\,,
\label{eq:maxS}
\eq
occurs at
\bq
k_0^2=\frac{1}{2}\frac{\left(2\,\alpha-1\right)^2\,s^2}{s-2}\,.
\label{eq:maxK}
\eq
Clearly, the instability reaches its maximum growth rate only when $s>2$.  Therefore, the waves corresponding to region III
in Fig.~(\ref{fig:F14}) exhibit this maximum growth rate.

{\bf {2. $C_0>0\, (D<0)$:}} The roots of $\sigma^2$ are complex, i.e. 
\bq
\sigma=\gamma-i\,\omega\,,
\eq
and waves are overstable. 

Consider the case $\alpha=1/2$ where Fig.~\ref{fig:F13}(a) shows $s=2$. Since $1-s+\alpha\,(1-\alpha)\,s^2=(1-s/2)^2$, $C_0$ remains positive and Eq.~(\ref{eq:disx})reduces to 
\bq
\sigma^2\pm i\,k^2\,\sigma + k^2\,\left(1-\frac{s}{2}\right)=0\,.
\eq
Thus,
\bq
\sigma=\pm \frac{i\,k^2}{2}\pm k\,\sqrt{\frac{s}{2}-1-\frac{1}{4}k^2}\,,
\eq
and the roots are overstable with
\bq
\omega=\frac{i\,k^2}{2}\,,\mbox{and}\quad \gamma=\frac{k}{2} \sqrt{2\,s-2-k^2}\,,
\eq
when $\frac{s}{2}-1-\frac{1}{4}\,k^2>0$ i.e. when $s>2$ and $k^2<2\,\left(s-2\right)$.
 
{\bf {3. $C_0>0\, (D>0)$:}} Two negative roots, $-\omega_1^2\,,-\omega_1^2$, or 
\bq
\sigma=\pm i\,\omega_1\,\quad\mbox{and}\quad \pm i\,\omega_2\,,
\eq
This is the only stable case.

\begin{figure}
\includegraphics[scale=0.34]{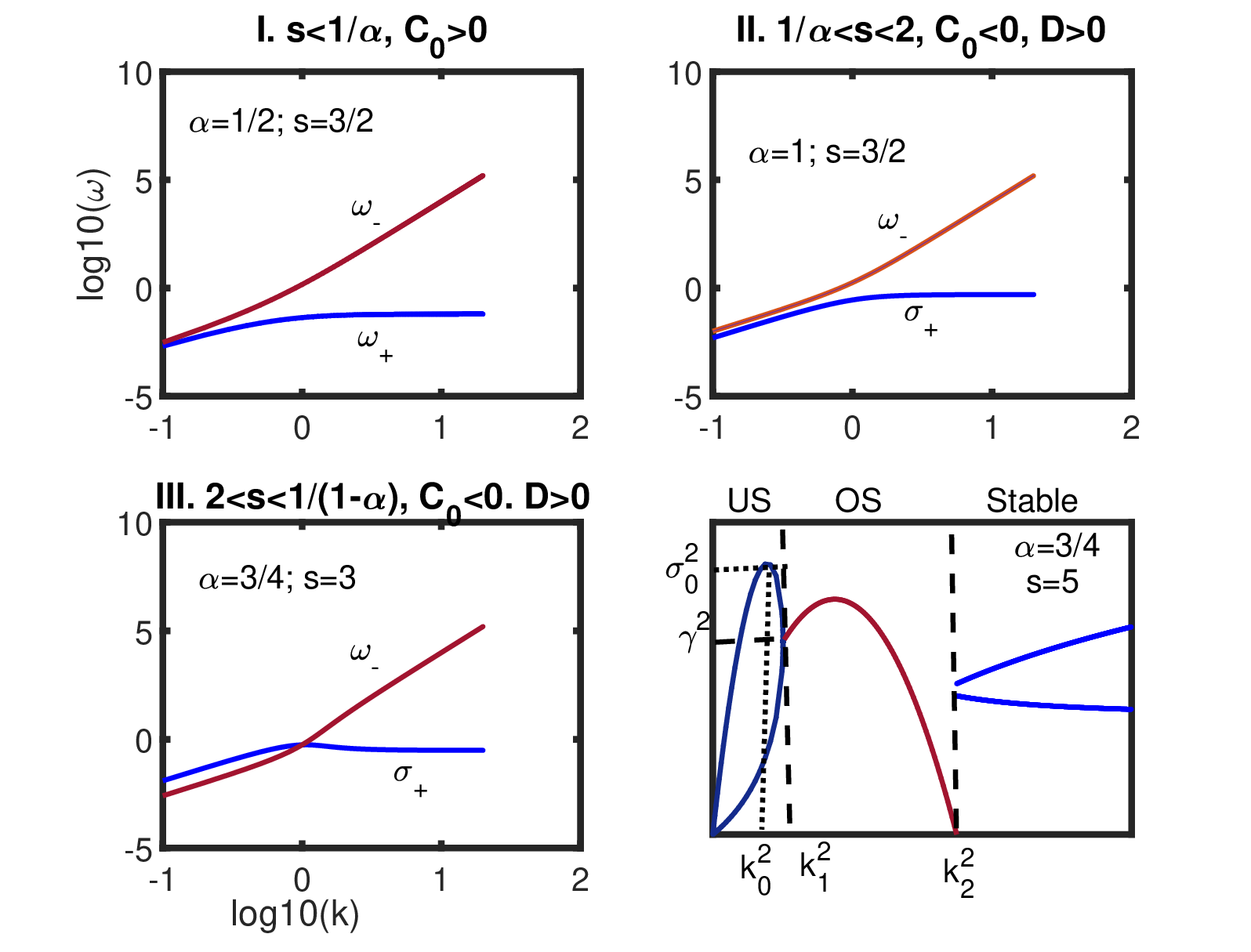}
\caption{Normal modes in the region I-IV of Fig.~(\ref{fig:F14}) is shown in the above figure\,.}
\label{fig:F15}  
\end{figure}

Now we analyze the stable, unstable and overstable regions, I-IV  of Fig.~(\ref{fig:F14}) in some detail.\\
{\bf {Stable region I: $s<1/\alpha\Rightarrow C_0>0$:}}\\
In this case, we have 
\bq
\omega_{-}^2=\frac{k^2}{2}\left(C_2+\sqrt{D}\right)
\approx \begin{cases}
\frac{k^2}{2}\Big[2-s+\left(2\,\alpha-1\right)\,s\Big]\,,\text{if}\,\,k\rightarrow 0\,\\
k^4\,,\text{if}\,\,k\rightarrow \infty\,,
\end{cases}
\eq
and 
\bq
\omega_{+}^2=\frac{k^4\,C_0}{\omega_{-}^2}
\approx \begin{cases}
\frac{k^2}{2}\Big[2-s-\left(2\,\alpha-1\right)\,s\Big]\,,\text{if}\,\,k\rightarrow 0\,\\ 
C_0\,,\text{if}\,\,k\rightarrow \infty\,.
\end{cases}
\eq
The frequencies $\omega_{-}$ and $\omega_{+}$ correspond to the shear-modified whistler and cyclotron waves, respectively. These normal modes are 
plotted as functions of $\log_{10}{k}$ in the top left panel of Fig.~\ref{fig:F15}, where $\alpha=1/2$ and $s=1.5$.

{\bf {Unstable region II: $1/\alpha<s<2\Rightarrow C_0<0\,,D>0$:}}\\
The roots of the Eq.~(\ref{eq:disx}) are
\bq
\sigma_{+}^2=\frac{k^2}{2}\left(\sqrt{D}-C_2\right)
\rightarrow \begin{cases}
\alpha\,s-1\,\\
\left(1-\alpha\right)\,s-1\,,
\end{cases}
\label{eq:rIIa}
\eq
where asymptotic expression in the above equation is written in the $k\rightarrow 0$ limit. 
The other root is
\bq
\omega_{-}^2=\frac{k^2}{2}\left(C_2+\sqrt{D}\right)\,.
\label{eq:rIIb}
\eq
These modes are shown in the top right corner of the Fig.~(\ref{fig:F15}) against $\log_{10}{k}$ for $\alpha=1$ and $s=1.5$.

{\bf {Unstable region III: $2<s<1/(1-\alpha)\,, C_0<0\,,D>0$:}}\\
Although the roots of Eq.~(\ref{eq:disx}) are same as in the previous case, the maximum growth rate, Eq.~(\ref{eq:maxS}) is attainable only in this region. 

{\bf {Unstable (US),  overstable (OS) and stable regions IV: $s>2\,, C_0>0$:}}\\
The unstable region IV is characterized by the roots given in Eq.~\ref{eq:rIIa}. An example of such a root for $\alpha=0.75$ and $s=5$ is plotted as a function of $k^2$ in the bottom right panel of Fig.~\ref{fig:F15}. The maximum growth rate $\sigma_0^2$ and its corresponding wavenumber $k_0^2$ are marked in the plot.

The overstable (OS), $\sigma^2=\left(\gamma\pm i\,\omega\right)^2$ root falls within $[k_1\,,k_2]$ interval. Here
\begin{eqnarray}
\omega^2=\frac{1}{4}k^2\,\left(k^2-k_1^2\right)\,,\nonumber\\
\gamma^2=\frac{1}{4}k^2\,\left(k_2^2-k^2\right)\,.
\end{eqnarray}

\begin{figure}
\includegraphics[scale=0.34]{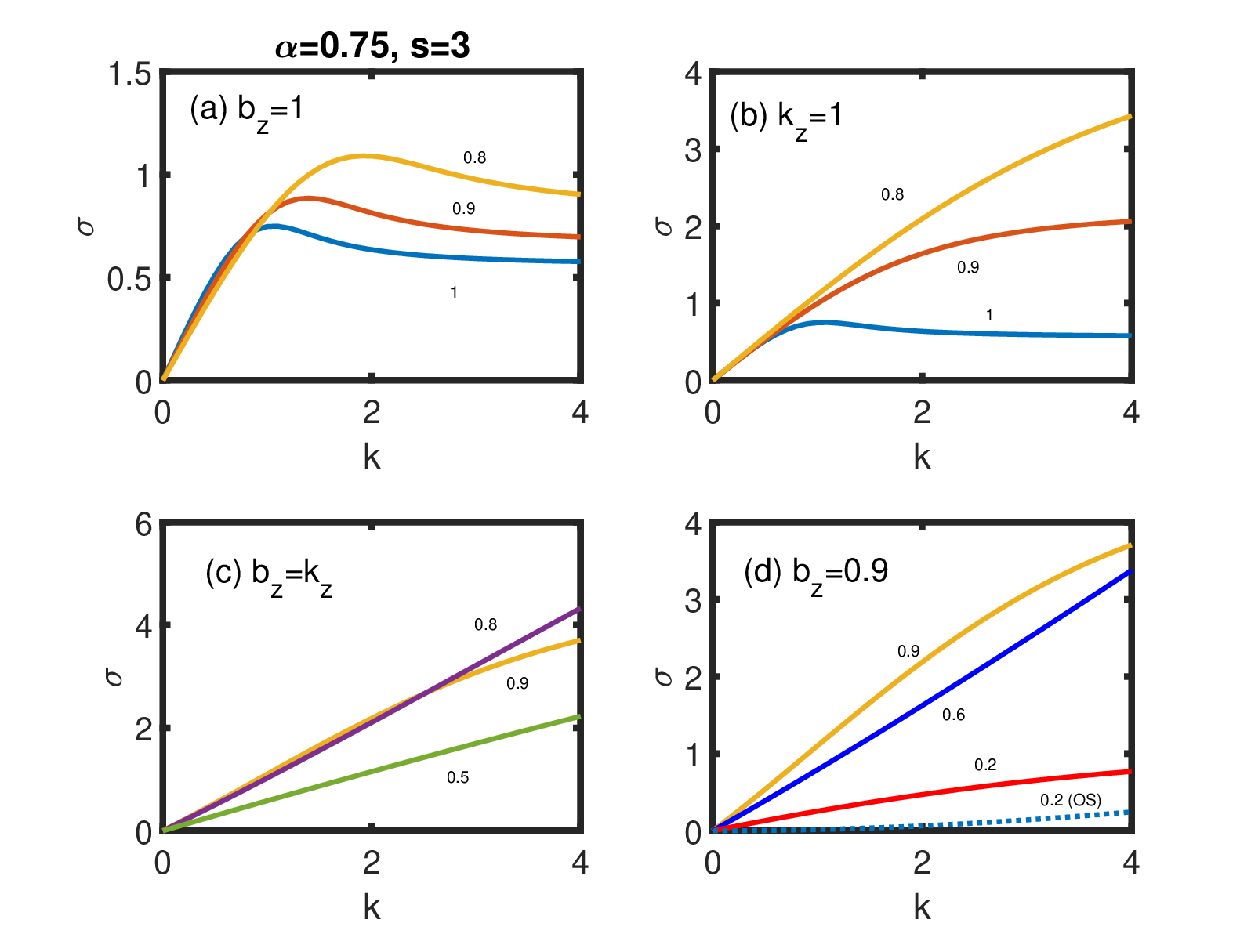}
\caption{Roots of the dispersion relation is plotted against $k$ for the fixed $\alpha=0.75\,,s=3$ in the above figure\,.}
\label{fig:F16}  
\end{figure}

We Note  from Eq.~(\ref{eq:disx}) that $\sigma^2=0$ when $k^2=0$. So $\sigma^2$ crosses $k^2$ axis at the origin. For small $k$, 
$D=\left(2\,\alpha-1\right)^2\,s^2$ and
\bq
\frac{\sigma^2}{k^2}=\begin{cases}
\alpha\,s-1\,,\\
\left(1-\alpha\right)\,s-1\,.
\end{cases}
\eq

In the transition region, $\alpha \gtrsim 1/2$ for both $50$ and $100,\mbox{G}$ fields [lower panel of Fig.~\ref{fig:F4}]. According to Fig.~\ref{fig:F14}, waves in this region may become unstable when $s\lesssim 2$. In the upper chromosphere, where $\alpha \approx 3/2$ for $50,\mbox{G}$ [lower panel of Fig.~\ref{fig:F4}], instability occurs only for $s<2$. To summarize, shear flow in the upper chromosphere-transition region can destabilize waves through gyroviscous momentum transport, with the value of $\alpha$ determining whether the waves exhibit overstability or instability.

Fig.\ref{fig:F13}(c) demonstrates that for $\alpha=1$, waves become unstable when $s>1$. This condition may be relevant to the upper chromosphere, where $\alpha \approx 1$ for magnetic fields of $B_0=100,\mbox{G}$ [Fig.\ref{fig:F4}]. For $\alpha>1$, instability can occur at lower shear values within the interval $s\in(1/\alpha, 1/(\alpha-1))$ [Fig.~\ref{fig:F13}(d)]. Thus, depending on the shear magnitude and the local value of $\alpha$, waves in different regions of the solar atmosphere can exhibit either overstability or instability.\\

{\bf{General case: $\mu\neq1$}}\\

For $\mu\neq 1$ and non-vertical magnetic field components, the dispersion relation Eq.\ref{eq:dr4} is solved numerically with $\alpha=3/4$ and $s=3$, with results shown in Fig.\ref{fig:F16}. Panels (a) and (b) demonstrate that the gyroviscous instability exhibits higher growth rates either when the magnetic field lacks an azimuthal component or when the wavevector has no radial component. In both panels, curves labeled '1' correspond to $b_z=\kz=1$. Panel (a) shows curves labeled by their $\kz$ values, while panel (b) uses $b_z$ values for labeling. The trends in panel (a) indicate that the instability growth rate increases as waves propagate more transversely, becoming increasingly magnetosonic in nature. Similarly, panel (b) shows that the growth rate increases with decreasing $b_z$.

In panel (c), where $b_z$ and $k_z$ are parallel, the instability exhibits higher growth rates when both field and wavevectors are nearly vertical ($\approx 0.8$, $0.9$). As $b_y$ and $\kx$ increase, making the wave more magnetosonic than \alfc in nature, the growth rate decreases (as seen in the curve labeled $0.5$). Panel (d) shows the behavior for fixed magnetic field configuration with varying $\kz$. At small $\kz$ values (e.g., the $0.2$ curve), the purely growing gyroviscous mode coexists with a slowly growing overstable mode [denoted by the dotted curve labeled 0.2 (OS)]. Thus, the magnetic field topology and wave characteristics determine whether shear flow with gyroviscosity leads to instability or overstability.

\section{Discussion}
The coronal thermal energy density is negligible compared to the photosphere, with densities of $\sim 10^8-10^9\,\cc$ versus $\sim 10^{17}\,\cc$ respectively. Despite this, the corona is $200-300$ times hotter than the photosphere. Since the second law of thermodynamics prevents heat flow from cooler to hotter regions, a mechanism must exist to heat the corona to million-degree temperatures.

Two primary heating mechanisms are wave heating and magnetic reconnection. In quiet solar regions with stable magnetic topology, waves are the primary energy transport mechanism from photosphere to corona. Active regions can utilize both wave heating and magnetic reconnection \citep{S17}.

Wave heating in the corona depends on partially ionized solar atmosphere microphysics. Strong collisional coupling causes these microphysical effects to manifest at macro scales, with spatial and temporal scales exceeding those in fully ionized plasmas. The ion-cyclotron frequency (Hall frequency) and ion-Larmor radius become functions of fractional ionization, resulting in lower frequencies and larger radii (PW06, PW08, PW22). This coupling leads to simultaneous effects of viscous momentum transport and non-ideal magnetic diffusion on wave propagation.

Fluid viscosity dampens waves and heats the fluid \citep{LL87}, with parallel ($\nu_0$) and perpendicular ($\nu_1\,,\nu_2$) viscosities having similar damping effects in partially ionized plasma. The relative importance of viscous versus ambipolar heating depends on the Prandtl number, varying between quiet and active solar regions. In quiet regions ($B=20\,\mbox{G}$), viscous damping dominates chromospheric and transition region heating, operating an order of magnitude faster than ambipolar diffusion. For stronger fields ($B=100\,\mbox{G}$), combined viscous and ambipolar damping exceeds pure ambipolar rates. In the transition region, declining neutral density reduces ambipolar diffusion, leaving parallel viscosity as the primary heating mechanism.

\begin{figure}
\includegraphics[scale=0.34]{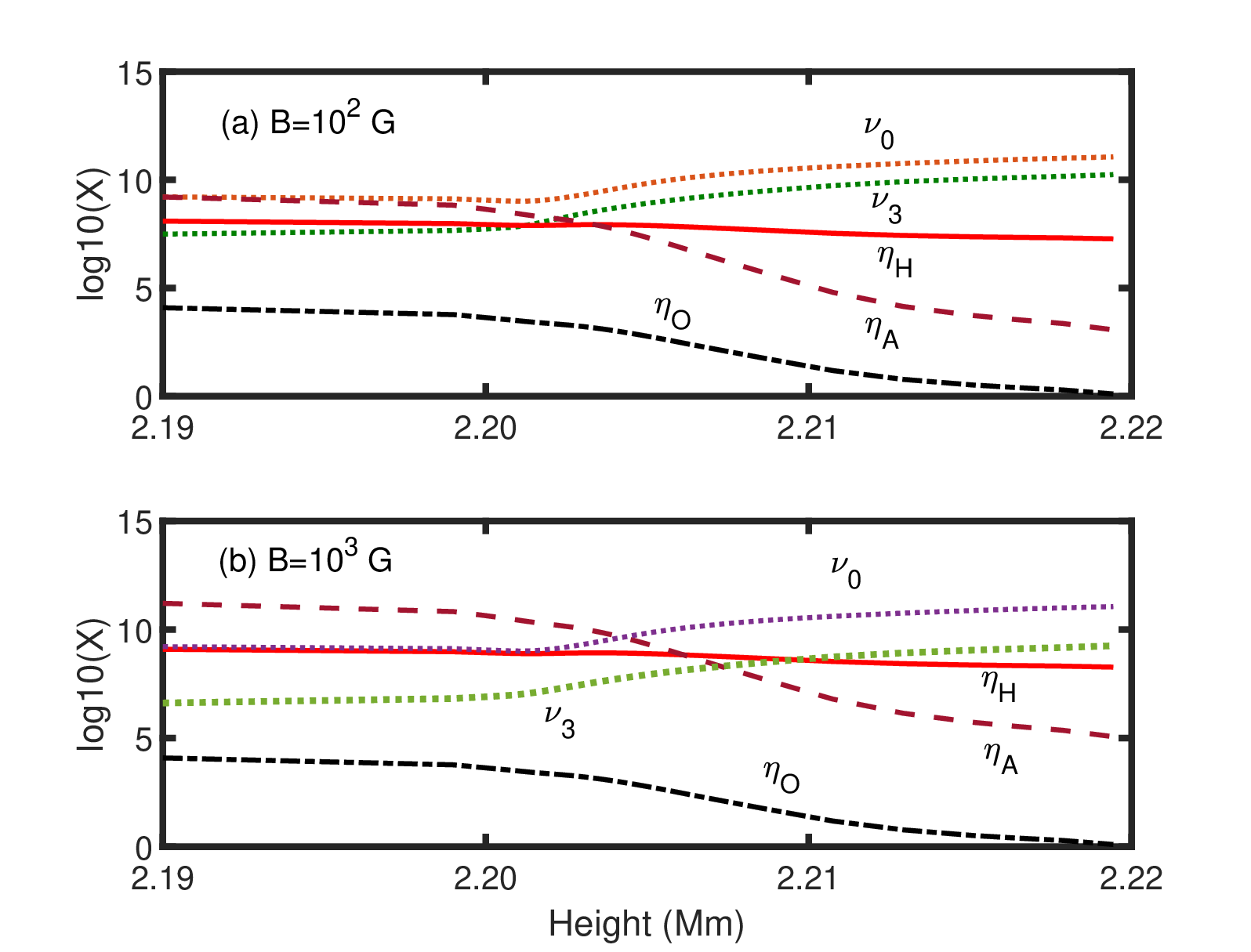}
\caption{Panel (a) shows magnetic diffusion coefficients, parallel viscosity, and FLR viscosities in the transition region for 
$B_0=100\,\mbox{G}$, while panel (b) shows these quantities for $B_0=1\,\mbox{kG}$.}
\label{fig:F17}  
\end{figure}

In Fig.\ref{fig:F17}(a), we plot the parallel ($\nu_0$) and gyro ($\nu_3$) viscosities alongside the magnetic diffusion coefficients ($\eta_H$, $\eta_A$, and $\eta_O$) in the transition region. For this analysis, we use $B_0=100\,\mbox{G}$ in panel (a) and $B_0=1\,\mbox{kG}$ in panel (b). The $\nu_1$ and $\nu_2$ profiles are omitted as they closely follow the $\nu_0$ profile [Fig.\ref{fig:F1})].

For $B_0=100\,\mbox{G}$, parallel viscosity and ambipolar diffusion remain comparable up to approximately $2.20\,\mbox{Mm}$, beyond which parallel viscosity becomes the dominant mechanism. This indicates that wave heating in the upper transition region is primarily driven by parallel viscosity. In regions with stronger magnetic fields (e.g., $B_0\gtrsim\mbox{1\,kG}$), the heating dynamics become more complex. As shown in Fig.~\ref{fig:F17}(b), ambipolar diffusion dominates up to $2.205\,\mbox{Mm}$, after which parallel viscosity becomes predominant. This spatial distribution of heating mechanisms—with ambipolar heating prevailing in the chromosphere and viscous heating dominating the transition region—suggests that models considering only one mechanism substantially underestimate the total heating rate.

The MHD wave-generated heat flux exhibits strong dependence on magnetic field strength. In quiet solar regions ($B_0\lesssim 100\,\mbox{G}$), the generated energy flux of $\sim 10^8\,\mbox{ergs}\,\mbox{cm}^{-2}\,\mbox{s}^{-1}$ is sufficient to balance coronal radiative losses. In sunspot regions characterized by strong magnetic fields ($B_0 \gtrsim 1\,\mbox{kG}$), even relatively small magnetic fluctuations ($\delta B = 0.1\B_0$) produce chromospheric heat flux that exceeds the required radiative losses by several orders of magnitude. These results indicate that efficient coronal heating requires either magnetic field strengths or wave amplitudes to exceed certain threshold values in both quiet and active regions.

Viscosities play dual roles in wave dynamics, affecting both wave damping and magnetic diffusion-driven instabilities in shear flows (PW13). The behavior varies by atmospheric layer:\\
(I) Photospheric Region:
In the photosphere, where viscosity is isotropic ($\nu_0=\nu_1=\nu_2$), its effect is limited to wave damping without coupling to shear flow. Wave excitation from shear flow energy occurs exclusively through Hall and ambipolar diffusion mechanisms. However, viscosity constrains these instabilities to long wavelength regimes.\\
(II) Chromospheric Region:
In the chromosphere, even slight anisotropy between perpendicular viscosities ($\nu_1-\nu_2\sim 10^{-2}$) enables coupling between viscosity and shear flow, leading to wave instability due to differential damping. The interaction with different diffusion mechanisms produces distinct effects. Ohm and ambipolar diffusion create two distinct wavelength regimes, (a) purely growing modes at short wavelengths and (b) overstable modes at long wavelengths. Hall diffusion, in contrast, restricts viscous instability to longer wavelengths with reduced growth rates, without inducing wavelength regime splitting

While parallel viscosity ($\nu_0$) typically acts as a wave damping mechanism, its interplay with Hall diffusion can trigger viscous-Hall instability under specific shear gradients (opposite to those driving Hall instability). This viscous-Hall instability reaches peak growth rates for nearly vertical fields and field-aligned wavevectors (almost \alfc waves), with growth rates scaling with the Hall-to-viscosity ratio ($\eta_H/\nu_0$).

In the lower and middle chromosphere ($0.5$-$1.2,\mbox{Mm}$), where both Hall diffusion and viscosity coexist with varying strengths, their competition establishes wavelength-dependent stability regimes. Hall diffusion dominates in this region, enabling short-wavelength instabilities. Above $1.2,\mbox{Mm}$, viscosity becomes dominant, suppressing short-wavelength fluctuations and constraining viscous instability to longer wavelengths.

The ratio of gyroviscosities ($\alpha=\nu_3/\nu_4$) in the transition region varies between $3/2$ and $1/2$ for magnetic field strengths of $50$ and $100\,\mbox{G}$ respectively [Fig.~\ref{fig:F4}]. In plasmas with purely vertical magnetic fields, gyroviscosity ($\alpha=1/2$) destabilizes the \alf wave (PW22, PW23). Our analysis reveals that wave stability—whether stable, overstable, or unstable—is determined by the combined effects of $\alpha$ and shear $s$. Specifically, for $\alpha \in [1/2,1]$, waves become unstable when shear lies in the range $1/\alpha < s < 1/\left(1-\alpha\right)$, with maximum growth rates achieved for $s > 2$. These conditions suggest that gyroviscous instability likely develops in the upper chromosphere-transition region.

The gyroviscous instability requires only non-zero vertical magnetic field and parallel wavevector (PW22), independent of the topological switch $g$. However, the magnetic field topology and wave characteristics (\alfc or magnetosonic) determine the stability behavior in shear flows. The growth rates exhibit specific dependencies:\\
Growth rates increase when either:\\
(i)The magnetic field lacks an azimuthal component, or\\
(ii)The wavevector lacks a radial component\\
indicating faster growth for predominantly magnetosonic waves.

For parallel $b_z$ and $k_z$, growth rates peak when both field and wavevectors are nearly vertical, favouring \alfc waves.
For magnetic fields with dominant vertical and small azimuthal components, the instability becomes overstable when wavevectors are predominantly radial with small vertical components.

The ultimate stability state (overstable or unstable) depends on both the shear gradient $s$ and topological factor $g$. These conditions suggest that gyroviscous instability likely destabilizes whistler and ion-cyclotron waves in the transition region.

The presence of viscosities in shear flows can generate turbulent plasma heating. Observations and numerical simulations reveal ubiquitous flow gradients throughout the photospheric-chromospheric plasma \citep{B08, At09, W09, Z93, SN98}, suggesting that parallel, perpendicular, and gyroviscosity may excite low-frequency turbulence at various altitudes.
While observed chromospheric swirls show relatively low vorticities ($\sim 6\times 10^{-3}\,\persec$, corresponding to $35$-minute rotation periods) \citep{B10}, these values are resolution-limited. Numerical simulations indicate significantly higher vorticities ($\sim 0.1$-$0.2\,\persec$) in the photosphere-lower chromosphere \citep{SN98}. At these higher values, viscous waves can become unstable within approximately one minute ($s = 0.2\,\persec$), making viscous instabilities likely in the chromosphere and transition region.
\section{summary}
The viscous and diffusive scales in partially ionized plasma vary with fractional ionization and magnetic field strength, causing different viscosity mechanisms to dominate at different atmospheric heights. The interplay between viscosities and magnetic diffusivities drives heating and turbulence throughout the solar atmosphere, with distinct regimes:\\
{\bf Photosphere-Chromosphere:}\\
Parallel and perpendicular viscosities dominate over gyroviscosity.\\
These viscosities show weak dependence on ion Hall parameter ($\beta_i$) and normalized Larmor radius ($k,R_L^*$).\\
Ohm diffusion exceeds viscous effects in the photosphere.\\
In middle/upper chromosphere, dominance of Hall-ambipolar diffusion versus viscous momentum transport depends on field strength.\\
{\bf Upper Chromosphere and Transition Region:}\\
Gyroviscosity becomes the primary viscosity mechanism.\\
Gyroviscous momentum transport dominates in the transition region.

Here's a summary of key findings.

1. Viscosity in the Photosphere-Chromosphere: Parallel and perpendicular viscosities are comparable in magnitude.
Perpendicular viscosities ($\nu_1$\, $\nu_2$) show weak ion-Hall parameter ($\beta_i$) dependence. 
Small anisotropy ($\sim 10^{-2}$) emerges in chromospheric viscous tensor.

2. Field-Strength Dependent Transport: For Weak fields ($B_0\lesssim 100\,\mbox{G}$) viscous transport dominates in chromosphere
For strong fields, viscous transport dominates in transition region while magnetic diffusion prevails in photosphere and lower/middle chromosphere.

3. Heating Mechanisms: In quiet regions (weak field), primarily viscous wave damping. 
In active regions ($B_0\gtrsim \mbox{1\,kG}$), combined ambipolar and viscous damping.
Upper chromosphere/transition region, predominantly viscous heating.

4. Isotropic Viscosity Effects: Inhibits Hall and ambipolar instabilities;
Introduces diffusive instability cutoff under shear flow conditions.
 
5. Anisotropic Viscosity Effects: Small $\nu_1$-$\nu_2$ differences enable all-wavelength instability;
Magnetic diffusion establishes viscous instability cutoff

6. Viscous-Hall Instability: Requires non-zero topological switch ($g=-\kx,\kz,b_y,b_z$);
Emerges from parallel viscosity-Hall diffusion interaction.

7. Gyroviscous Effects: Instability onset depends on $\nu_3/\nu_4$ ratio and shear ($s$);
Growth rates determined by magnetic topology and wavevector orientation.

\section*{Data Availability Statement}

Data sharing not applicable-no new data generated.

\appendix
\section[]{Linearized viscous tensor components}
As the spatial scale over which flow and field generation occurs, is much smaller than the typical tube diameter,  we shall approximate the cylindrical tube by a planer sheet and work in the Cartesian coordinate system where $x\,,y\,,z$ represents  radial, azimuthal and vertical directions locally. We shall assume initial homogeneous state with an azimuthal shear flow $\v = s\,x\,\bmath{y}$. Here $s\equiv {v_0}^{\prime}$. The magnetic field in the intergranular lanes at the network boundaries is clumped into elements or flux tubes that are generally vertical \citep{M97, H09} but highly inclined fields have also been reported in the literature \citep{S87a}.  The internetwork magnetic elements have predominantly horizontal field \citep{H09, S12}. Therefore, we shall assume uniform background field with both azimuthal as well as vertical component, $\B = (0, B_y, B_z)$.

For the magnetic field geometry  $\bb=\bm{y}\,b_y+\bm{z}\,b_z$ we have 
\bq
\bb\bb-\frac{1}{3}\tenq{\bf{I}}=
\left(
\begin{array}{ccc} -1/3 &  0 & 0\\
                    0 & \by2-1/3 & b_y\,b_z\\
                      0    &  b_y\,b_z     & \bz2-1/3    
  \end{array}
\right)\,.
\label{eq:fd1}
\eq
For the linear shear $v_y=s\,x$ where $s=\vp=const.$ only $W_{xy}=s$ is non-zero. Thus
\bq
\delta \tenq{W}_{0}=\frac{3}{2} \delta\left(\bb\cdot\tenq{W}\cdot\bb\right)\left(\bb\bb-\frac{1}{3}\tenq{\bf{I}}\right)\,, 
\eq
since $\left(\bb\cdot\tenq{W}\cdot\bb\right)\delta\left(\bb\bb-\frac{1}{3}\tenq{\bf{I}}\right)=0\,.$ For the assumed field geometry in the present work we have
\begin{eqnarray}
\left(\dbb\cdot \tenq{W}\cdot\bb \right)+\left(\bb \cdot \tenq{W} \cdot\dbb\right) = 2\,s\,b_y\dBxh\,,
\nonumber\\
\left(\bb\cdot \delta \tenq{W}\cdot\bb \right)=2\,b_y\,b_z\delta W_{yz} + b_z^2 \delta W_{zz}\,.
\end{eqnarray}
Linearising Eqs.~(\ref{eq:cont}) and Fourier analyzing with $k = (k_x , 0 , k_z)$, and $\delta \v=(\delta v_x\,,\delta v_y\,, \delta v_z)$ in Boussinesq approximation, we have
\bq
k_z\,\delta \vz+k_x\delta \vx = 0\,.
\label{eq:cd}
\eq
Further 
\begin{eqnarray}
 \delta W_{xx}=-\delta W_{zz}=2\,i\,k_x\,\dvx\,,\delta W_{yy}=0\,,
 \nonumber\\
 \delta W_{xy}=i\,k_x\,\dvy\,,\delta W_{xz}=i\,f\,\dvx\,,\delta W_{zy}=i\,k_z\,\dvy\,,
\end{eqnarray}
where
\bq
f=\frac{k_z^2-k_x^2}{k_z}
\eq
Thus,
\bq
\delta \tenq{W}_{0}=A\,\left(
\begin{array}{ccc} -1 &  0 & 0\\
                    0 & 3\by2-1  & 3\,b_y\,b_z\\
                      0    &3\,b_y\,b_z& 3\,b_z^2-1    
  \end{array}
\right)\,.
\eq
where 
\bq
A=s\,b_y\,\dBxh + i\,\bz2\,\left(R\,k_z\,\dvy-k_x\,\dvx\right)\,.
\eq
Here $R=\frac{b_y}{b_z}$.

Similarly, 
\bq{}
 \delta\tenq{W}_{1}=
\left(
\begin{array}{ccc} c_{11} & c_{12} & c_{13}\\
                    c_{12} & c_{22} & c_{23}\\
                      c_{13}    & c_{23}   & c_{33}     
  \end{array}
\right)\,.
\eq{}
Here
\begin{eqnarray}
c_{11}=i\Big[\left(1+\by2\right)k_x\dvx+R\bz2\,k_z\dvy\Big]-s\,b_y\dBxh\,,
\nonumber\\
c_{22}=-i\bz2\Big[\left(1+\by2\right)k_x\dvx+R\bz2\,k_z\dvy\Big]-s\,b_y\bz2\dBxh\,,
\nonumber\\
c_{33}=-i\by2\Big[\left(1+\by2\right)k_x\dvx+R\bz2\,k_z\dvy\Big]+s\,b_y\left(1+\bz2\right)\dBxh\,,
\nonumber\\
\end{eqnarray}
and
\begin{eqnarray}
c_{12}=i\bz2\left(k_x\dvy-R\,f\dvx\right)-2\,s\,b_y\dByh\,,
\nonumber\\
c_{13}=-i\,R\bz2\left(k_x\dvy-R\,f\dvx\right)-s\,\left(b_y\dBzh+b_z\dByh\right)\,,
\nonumber\\
c_{23}=i\,R\bz2\Big[R\bz2\,k_z\dvy+\left(1+\by2\right)\,k_x\dvx\Big]-s\,b_z^3\dBxh\,.
\end{eqnarray}
Also
\bq{}
 \delta\tenq{W}_{2}=
\left(
\begin{array}{ccc} 2\,d_{11} & d_{12} & d_{13}\\
                    d_{12} &2\,d_{22} & d_{23}+d_{32}\\
                      d_{13}   & d_{23}+d_{32}   &2\,d_{33}     
  \end{array}
\right)\,,
\eq{}
with
\begin{eqnarray}
d_{11}=s\,b_y\,\dBxh\,,
\nonumber\\
d_{12}=i\,R\bz2\left(f\dvx+R\,k_x\dvy\right)+2\,s\,b_y\dByh\,,
\nonumber\\
d_{13}=i\,\bz2\left(f\dvx+R\,k_x\dvy\right)
+s\left(b_z\dByh-\frac{k_x}{k_z}b_y\dBxh\right)\,,
\nonumber\\
d_{22}=i\,R\,b_z^4\left(2\,R\,k_x\dvx+R_1k_z\dvy\right)
\nonumber
+s\,R\,b_z^2R_1\dBxh\,,
\nonumber\\
d_{32}/R^2=-i\,b_z^4\left(2\,R\,k_x\dvx+R_1k_z\dvy\right)-2\,sb_z^3\dBxh\,,
\nonumber\\
d_{23}=d_{22}/R\,,d_{33}=d_{32}/R\,,
\nonumber\\
\end{eqnarray}
with $R_1=1-R^2$.
Defining
\begin{eqnarray}
a_{11}=i\,b_z\,\left(R\,f\,\dvx-k_x\dvy\right)\,,
\nonumber\\
a_{12}=i\,b_y\left(\left(1+Q\right)k_z\,\dvy-2\,N\,b_y\,b_z\,k_x\dvx\right)\,,
\nonumber\\
a_{13}=-i\,b_z\left(\left(1-Q\right)k_z\,\dvy+2\,R\,Q_1\,k_x\dvx\right)\,,
\nonumber\\
a_{21}=2\,i\,b_z\,k_x\dvx\,,
\nonumber\\
a_{22}=i\,b_z\left(Q_2\,k_x\,\dvy+N\,R\,\bz2\,f\,\dvx\right)\,,
\nonumber\\
a_{23}=i\,b_z\left(Q_1\,f\,\dvx+N\,R\,\bz2\,k_x\,\dvy\right)]\,,
\nonumber\\
a_{31}=s\dBxh-2\,i\,b_y\,k_x\,\dvx\,,
\nonumber\\
a_{32}=-i\,b_y\left(Q_2\,k_x\,\dvy+N\,R\,\bz2\,f\,\dvx\right)-sQ_2\dByh \,,
\nonumber\\
a_{33}=-i\,b_y\left(Q_1\,f\,\dvx+N\,R\,\bz2\,k_x\,\dvy\right)-N\,s\,b_y\,b_z\dByh\,,
\end{eqnarray}
with $Q/N=b_y^2-b_z^2$, $Q_1=1-b_z^2$, $Q_2=1-b_y^2$,  the linearized gyroviscous tensor becomes
\bq
  \left(\delta \tenq{W}_{3}+2\,\alpha\,\delta \tenq{W}_{4}\right)+\left(\right)^{\mbox{T}}=\frac{1}{2}
  \left(
\begin{array}{ccc} S_{11} &  S_{12} & S_{13}\\
                    S_{12} & S_{22} & S_{23}\\
                      S_{13}   & S_{23}     & S_{33}    
  \end{array}
\right)\,.
\eq
Note that the $\left(\right)^{\mbox{T}}$ is the transpose of the matrix $\left(\delta \tenq{W}_{3}+2\,\alpha\,\delta \tenq{W}_{4}\right)$ in the above expression. Here
\begin{eqnarray}
 S_{11}=2\,a_{11}\,,S_{12}=a_{12}+a_{21}\,,S_{13}=a_{13}+a_{31}-N\,s\,\dBxh\,,
 \nonumber\\
 S_{22}=2\,a_{22}+4\,N\,s\,b_y\,b_z\,\dByh\,,
 \nonumber\\
 S_{23}=a_{23}+a_{32}+N\,s\,\Big[\left(1-3\by2\right)\dByh+\frac{g}{\kz^2} \dBxh\Big]
 \nonumber\\
S_{33}=2\,a_{33}+2\,N\,s\,b_y\left(\frac{k_x\,b_y}{\kz}\dBxh-b_z\dByh\right)\,.
 \end{eqnarray}

\section[]{Waves in a homogeneous medium} 

The linearized and Fourier analyzed momentum equation becomes
\begin{eqnarray}
\sigma \,\dvx&=&- \,i\,k_x\,\frac{\delta p}{\rho}- \frac{1}{\rho}\left(\div \delta \Pi\right)_{x} + M_x\,, \nonumber\\
\sigma\,\dvy + s\,\dvx &=& -\frac{1}{\rho}\left(\div \delta \Pi\right)_{y}+M_y\,,\nonumber\\
\sigma\,\dvz &=& - i\,k_z\,\frac{\delta p}{\rho} -\frac{1}{\rho}\left(\div \delta \Pi\right)_{z}+ M_z\,.
\label{eq:cfm}
\end{eqnarray}
Here
\begin{eqnarray}
M_x&=& i\,\va^2\Big[\left(\k\cdot\Bh\right)\dBxh-\left(\Bh\cdot\dBh\right)k_x\Big]\,,
\nonumber\\
M_y&=& i\,\va^2\left(\k\cdot\Bh\right)\dByh\,,
\nonumber\\
M_z&=& i\,\va^2\Big[\left(\k\cdot\Bh\right)\dBzh-\left(\Bh\cdot\dBh\right)k_z\Big]\,,
\end{eqnarray}
The components of $\div \Pi\,$ are
 \begin{eqnarray}
\left(\div \Pi\right)_{x}=\del_x\Pi_{xx}+\del_y\Pi_{xy}+\del_z\Pi_{xz}\,, 
\nonumber\\
\left(\div \Pi\right)_{y}=\del_x\Pi_{xy}+\del_y\Pi_{yy}+\del_z\Pi_{yz}\,,
\nonumber\\
\left(\div \Pi\right)_{z}=\del_x\Pi_{xz}+\del_y\Pi_{yz}+\del_z\Pi_{zz}\,,
\end{eqnarray}
Here $\del_r=\partial/\partial r$. 
The various components of the linearized symmetric tensor $\delta \tenq{\Pi}$ becomes
\begin{eqnarray}
 \rho\,\delta \Pi_{xx}&=&\nu_0\,A-\nu_1\,c_{11}-2\,\nu_2\,d_{11}+\frac{\nu_3}{2}\,S_{11}\,,
 \nonumber\\
 \rho\,\delta \Pi_{yy}&=&\nu_0\,\left(1-3\,\by2\right)\,A-\nu_1\,\,c_{22}-2\,\nu_2\,d_{22}+\frac{\nu_3}{2}\,S_{22}\,, 
 \nonumber\\
 \rho\,\delta \Pi_{zz}&=&\frac{\nu_0}{2}\,\left(1-3\,\bz2\right)\,A-\nu_1\,c_{33}-2\,\nu_2\,d_{33}+\frac{\nu_3}{2}\,S_{33}\,,
 \nonumber\\
 \rho\,\delta \Pi_{xy}&=&-\nu_1\,c_{12}-\nu_2\,d_{12}+\frac{\nu_3}{2}\,S_{12}\,,
 \nonumber\\
 \rho\,\delta \Pi_{xz}&=&-\nu_1\,c_{13}-\nu_2\,d_{13}+\frac{\nu_3}{2}\,S_{13}\,,
 \nonumber\\
 \rho\,\delta \Pi_{yz}&=&-3\,\nu_0\,R\,\bz2\,A-\nu_1\,c_{23}-\nu_2\left(d_{23}+d_{32}\right)
 \nonumber\\
 {}&{}&+\frac{\nu_3}{2}S_{23}\,.
\label{eq:lpt}
\end{eqnarray}

From the  $x$ and $z$-components of the momentum equation (\ref{eq:cfm}), after making use of Eq.~(\ref{eq:cd}) we get
\bq
\delta p=-\kx^2\,\delta \Pi_{xx}-2\,\kx\,\kz\,\delta \Pi_{xx} - \kz^2\,\delta \Pi_{zz}-\rho\,\va^2\left(\Bh\cdot\dBh\right) 
\label{eq:pE}
\eq
Making use of the above Eq.~(\ref{eq:pE}) in the $x-$component of the momentum Eq.~(\ref{eq:cfm}) gives
\bq
\sigma \,\dvx = -\frac{i\,k_x\,\kz^2}{\rho}\bigg[\delta \Pi_{xx}-\delta \Pi_{zz}+ \frac{f}{k_x}\delta \Pi_{xz}\bigg] + i\,k\,\mu\,\va^2\, 
\dBxh\,.
\eq
Here
$$f= \frac{k_z^2-k_x^2}{k_z}\,.$$
and $\mu$ is given by Eq.~(\ref{eq:AFR}).
The $y-$component of the momentum equation is 
\bq
\sigma\,\dvy + s\,\dvx= -\frac{i}{\rho}\,\left(k_x\, \delta \Pi_{xy}+ k_z\,\delta \Pi_{yz}\right)+ M_y\,.
\eq
Making use of Eq.~(\ref{eq:lpt}) the components of $\delta \Pi$ can be expressed in terms of $\dvx\,,\dvy$. Thus the $x$ and $y$ component of the momentum equation becomes
\begin{eqnarray}
 \left(
\begin{array}{cc} \sigma+X_1 & X_2\\
                    
                 s+Y_1   & \sigma+Y_2    
  \end{array}
\right)\,\frac{\delta \v_{\perp}}{\va}
\nonumber\\
= i\,
\left(
\begin{array}{cc} \mu\,\omA-s_1\,S_{xx} & -s_1\,S_{xy}\\
                    
                 -s_1\,S_{yx}   & \mu\,\omA-s_1\,S_{yy}    
  \end{array}
\right)\,
\delta {\bb}_{\perp}
\,,
\label{eq:fdm}
\end{eqnarray}
Here $s_1=s/\omega_A$ and $\omega_A$ is the \alf frequency (Eq.~\ref{eq:AFR}). Defining
\bq
\Delta_0=\omega_0-\omega_2\,,\quad
\Delta_1=\omega_1-\omega_2\,,\quad F_1=1-\mu^2\,,
\eq
we may write
\begin{eqnarray}
X_1&=&\frac{3\,g^2}{R^2}\,\Delta_0+F_1\,\left(\by2+\mu^2\right)\,\Delta_1
+ \omega_2 
\,,
\nonumber\\
X_2&=&3\,\mu^2\,g\,\Delta_0-F_1\,\,g\,\Delta_1+\Big[F_1\,\left(\omega_3-\omega_4\right)+\mu^2\,\omega_4\Big]
\,\mu\,\kz\,,
\nonumber\\
Y_1&=&3\,\bz2\,g\,\Delta_0-F_1\,\frac{g}{\kz^2}\Delta_1
-\Big[F_1\,\left(\omega_3-\omega_4\right)+\mu^2\,\omega_4\Big]\,b_z\,,
\nonumber\\
Y_2&=&3\,\mu^2\,\by2\,\Delta_0+F_1\,\bz2\,\Delta_1 
+\omega_2\,,
\end{eqnarray}
and 
\begin{eqnarray}
S_{xx}=-3\,g\,\mu\,\Delta_0+\left(1+\mu^2\right)\,\kx\,b_y\,\Delta_1
\nonumber\\
+\Big[\left(1-\kx^2\,\bz2\right)\left(\omega_3-\omega_4\right)+\kx^2\,\bz2\,\omega_4\Big]\,\kz\,,
\nonumber\\
S_{xy}=\mu\,\left(\kz^2-\kx^2\right)\Delta_1+2\,\left(\omega_3-2\,\omega_4\right)\,\kz\,g\,,
\nonumber\\
S_{yx}=-3\,\mu\,\by2\,\Delta_0+\mu\,\bz2\,\Delta_1+
\left(2\,\omega_4-\omega_3\right)\,\frac{g}{\kz}\,,
\nonumber\\
S_{yy}=2\,\kx\,b_y\,\Delta_1
-\Big[\left(\bz2-\by2\right)\left(\omega_3-\omega_4\right)+2\,\by2\omega_4\Big]\,\kz\,.
\end{eqnarray}
Recall that $\omega_j=k^2\,\nu_j$ is the viscous frequency.  Eq.~(\ref{eq:fdm}) is Eq.~(1) of [Pandey \& Wardle (2023); Hereafter PW23] for $\mu=1$ or $b_z=\kz=1$ and $\alpha=1$.

The linearised and Fourier transformed induction equation becomes
\bq
 \left(
\begin{array}{cc} \sigma+\omega_{xx} & \omega_{xy}\\
                    
                 \omega_{yx}-s   & \sigma+\omega_{yy}    
  \end{array}
\right)\,\delta {\bb}_{\perp}
= i\,\mu\,\omA\,\frac{\delta \v_{\perp}}{\va} 
\,,
\label{eq:ind1}
\eq
Here 
\begin{eqnarray}
\omega_{xx}&=&k^2\,\left(\eta_O + b_z^2\,\eta_A\right)\,,
\nonumber\\
\omega_{xy}&=&k^2\,\left(H\,\eta_H + g\,\eta_A\right)\,,
\nonumber\\
\omega_{yx}&=&k^2\,\left(g\,\eta_A - \,\eta_H\right)/\kz^2\,,
\nonumber\\
\omega_{yy}&=&k^2\,\big[\eta_O + \left(1 -\kx^2\,b_z^2\right)\,\eta_A\big]\,,
\label{eq:ind_cof}
\end{eqnarray}
 are the various component of diffusivity tensor \cite{PW12}. Here the topological switch is
\bq
g = - \kx\,\kz\,b_y\,b_z\,,
\eq 
and 
helicity is
\bq
H = \mu\,\kz \equiv b_z\,\kz^2\,.
\label{eq:hct}
\eq

From equations (\ref{eq:fdm}) and  (\ref{eq:ind1}) we derive the following linear dispersion relation 
\begin{eqnarray}
\sigma^4+\left(C_3+E_3\right)\sigma^3+\left(C_2+E_2\right)\sigma^2+\left(C_1+E_1\right)\sigma
\nonumber\\
+\left(C_0+E_0\right)=0\,,
\label{eq:DRM}
\end{eqnarray}
where after defining
\begin{eqnarray}
A_1&=&X_1\,Y_2-X_2\,Y_1-s\,X_2+\left(\mu\,\omega_A\right)^2\,,
\nonumber\\
T_1&=&Y_1\,S_{xy}-X_1\,S_{yy}\,,
\nonumber\\
T_2&=&X_2\,S_{yx}-Y_2\,S_{xx}\,,
\nonumber\\
Z_1&=&\mu\,\left(X_2\,S_{yy}-Y_2\,S_{xy}\right)\,,
\nonumber\\
Z_2&=&\mu^2\,\left(S_{xx}\,S_{yy}-S_{xy}\,S_{yx}\right)\,,
\end{eqnarray}
we have
\begin{eqnarray}
C_3&=&X_1+Y_2\,,
\nonumber\\
C_2&=&A_1+\left(\mu\,\omega_A\right)^2-G_2(s)\,,
\nonumber\\
C_1&=&\left(\mu\,\omega_A\right)^2\,C_3+G_1(s)\,,
\nonumber\\
C_0&=&\left(\left(\mu\,\omega_A\right)^2-G_2(s)\right)\left(\mu\,\omega_A\right)^2+G_0(s)\,.
\label{eq:CoE1}
\end{eqnarray}
Here 
\begin{eqnarray}
G_2(s)&=&\mu\,s\left(S_{xx}+S_{yy}\right)\,,
\nonumber\\
G_1(s)&=&\mu\,s\,\left(T_1+T_2\right)\,,
\nonumber\\
G_0(s)&=&s^2\,\left(Z_1+Z_2\right)-s\,X_2\,\left(\mu\,\omega_A\right)^2\,.
\label{eq:GoE1}
\end{eqnarray}
The diffusion and the mixed (diffusion plus viscous) terms are contained in the various $E$ coefficients
which are
\begin{eqnarray}
E_0&=&b_{xx}\,\omega_{xx}+b_{xy}\,\omega_{xy}+b_{yx}\,\omega_{yx}+b_{yy}\,\omega_{yy}\,,
\nonumber\\
E_1&=&q_{xx}\,\omega_{xx}+q_{xy}\,\omega_{xy}+q_{yx}\,\omega_{yx}+q_{yy}\,\omega_{yy}\,,
\nonumber\\
E_3&=&\omega_{xx}+\omega_{yy}\,,
\nonumber\\
E_2&=&\left(X_1+Y_2\right)E_3+\omega_{xx}\omega_{yy}+s\omega_{xy}-\omega_{xy}\omega_{yx}\,.
\nonumber\\
\label{eq:deE}
\end{eqnarray}
Here
\begin{eqnarray}
b_{xx}&=&\left(\mu\,\omA\right)^2\,\left(X_1-\omega_{yy}\right)+A_1\,\omega_{yy}+\mu\,s\,\left(T_1+s\,S_{xy}\right)\,,
\nonumber\\
b_{xy}&=&\left(\left(\mu\,\omA\right)^2-s\,X_2-\mu\,s\,S_{xx}\right)\left(Y_1+s\right)
+s\,X_1\left(Y_2+\mu\,S_{yx}\right)\,,
\nonumber\\
b_{yx}&=&\left(X_2+\omega_{xy}\right)\left(\mu\,\omA \right)^2-A_1\,\omega_{xy}+\mu\,s\left(Y_2\,S_{xy}-X_2\,S_{yy}\right)\,,
\nonumber\\
b_{yy}&=&\left(\mu\,\omA\right)^2\,Y_2+\mu\,s\,T_2\,,
\end{eqnarray}
and
\begin{eqnarray}
q_{xx}&=&A_1-\mu\,s\,S_{yy}+C_3\,\omega_{yy}\,,
\nonumber\\
q_{yy}&=&A_1-\mu\,s\,S_{xx}\,,
\nonumber\\
q_{xy}&=&s\,\left(C_3+\mu\,S_{yx}\right)\,,
\nonumber\\
q_{yx}&=&-C_3\,\omega_{xy}+\mu\,s\,S_{xy}\,.
\end{eqnarray}

\end{document}